\documentstyle[12pt,epsfig]{article}
\def\shiftdown#1{#1\llap{\lower.04ex\hbox{#1}}}

\begin{document}
\vspace*{0.3cm}

\begin{center}

{\bf \large
Ward-Takahashi identity, soft photon theorem and
the magnetic moment of the $\Delta$ resonance.}
{\footnotemark}

\footnotetext{ 
Supported by the "Deutsche Forschungsgemeinschaft" under contract
GRK683
}

\end{center}


\vspace{5mm}

\noindent{
{\large \bf A.\ I.\ Machavariani$^a$ $^b$ $^c$ and Amand Faessler $^a$ } 

}

\vspace{5mm}

\noindent{\small

{  \rm $^a$ 
Institute\ f\"ur\ Theoretische\ Physik\ der\ Univesit\"at\
 T\"ubingen,\newline T\"ubingen\ D-72076, \ Germany}\\

{\rm $^b$ Joint\ Institute\ for\ Nuclear\ Research,\ Dubna,\ Moscow\
region\ 141980,\ Russia}\\

{\rm $^c$ High Energy Physics Institute of Tbilisi State 
University,
University str.  9,  Tbilisi 380086, Georgia  }\\




}

\vspace{0.5cm}


\medskip
\begin{abstract}
{\bf

Starting from the Ward-Takahashi identity for the radiative $\pi N$ 
scattering amplitude a generalization of the soft photon theorem approach is 
obtained for an arbitrary energy of an emitted photon.
The external particle radiation part of the $\pi N\to\gamma'\pi'N' $ 
amplitude is analytically reduced to the double $\Delta$ exchange amplitude
with the $\Delta\to\gamma'\Delta'$ vertex function.

We have shown, that the double $\Delta$ exchange amplitudes with 
internal $\Delta$ 
radiation is connected by current conservation with the corresponding part 
of the external particle radiation terms. Moreover according to
the current conservation the internal and external particle 
radiation terms with the $\Delta-\gamma'\Delta'$ vertex have a opposite sign 
i.e. they must cancel each other.
 Therefore we have a screening of the internal double $\Delta$ exchange 
diagram with the $\Delta-\gamma' \Delta'$ vertex
by the external particle radiation.
This enables us to obtain a model independent estimation of
the dipole magnetic moment of 
$\Delta^+$ and $\Delta^{++}$ resonances $\mu_{\Delta}$ through the 
anomalous magnetic moment
of the proton $\mu_p$ as $\mu_{\Delta^+}={ {M_{\Delta}}\over {m_p}} \mu_p$ 
and $\mu_{\Delta^{++}}={3\over 2}\mu_{\Delta^+}$
in agreement with the values obtained from the fit of the 
experimental cross section of the $\pi^+ p\to\gamma'\pi^+ p$ 
reaction. }

\end{abstract}


\newpage

\begin{center}
                  {\bf 1. INTRODUCTION}
\end{center}
\medskip

The bremsstrahlung reactions in the low and intermediate energy region
(up to 1GeV) are often investigated via the low energy photon theorem
\cite{Low}-\cite{PasVan2}. This approach enables 
us to calculate the bremsstrahlung 
amplitude in an expansion in powers of the small momentum $k'$ 
of the emitted photon
$A=a+b/k'+c{k'}^2+...$, where the first two terms can be reproduced exactly
with the corresponding non-radiative amplitudes. 
A starting point for the low energy theorems for the bremsstrahlung reactions
are the external particle radiation diagrams (Fig. 1) which determine the 
infrared behavior of this reaction. Using the current conservation 
condition one obtains a model independent connection between  
the external and internal particle radiation terms.
Corresponding internal particle radiation  terms allow to extract the
values of the resonances from the experimental data. However the prescription 
for the construction of the bremsstrahlung amplitude in the low energy limit 
$k'\to 0$ is not unique
and there  ambiguities appear from the soft photon approach \cite{Ding}.

In this paper we study the pion-nucleon bremsstrahlung based on the 
the Ward-Takahashi identity, because
 this identity generates
the current conservation for the 
on mass shell bremsstrahlung amplitude 
$k'_{\mu}A^{\mu}=0$ using 
the equal-time commutators between the photon current operator and
external particle field operators. On the other hand
 these equal-time commutator relations
follow from  charge conservation. 
In this way the current conservation for the 
full bremsstrahlung amplitude $k'_{\mu}A^{\mu}=0$ reduces to the
current conservation for the external particle radiation part
${\cal E}^{\mu}$, i.e.
$k'_{\mu}A^{\mu}=k'_{\mu}{\cal E}^{\mu}+B=0$, where $B$ is a
combination of the 
non-radiative off mass shell scattering amplitudes. 
This form of the current conservation condition is exactly the same as 
in the low energy theorems. 
But the present approach is not restricted  to  the low energy limit 
of the  final photon 
 and it can be applied for other reactions with  
electromagnetic interactions like pion photo-production, 
Compton scattering etc.
Current conservation in this approach provides  a 
model-independent connection
between  the external   ${\cal E}^{\mu}$ and internal 
${\cal I}^{\mu}$ particle radiation terms. Moreover, the mechanism of 
current conservation indicates, that the corresponding set of the
external and the internal particle
radiation terms have the opposite sign, i. e. they must cancel.
Thus in the considered reactions a screening of the 
internal particle radiation terms by the external particle radiation diagrams 
must be observed.

The suggested approach is applied to the radiative $\pi N$ scattering reaction
in the $\Delta$ resonance region. 
Using the projections on the spin $3/2$ amplitudes 
with the $\Delta$ exchange propagators,
from the external particle radiation
part of the $\pi N$ bremsstrahlung amplitude  the 
double $\Delta$ exchange part with the $\Delta-\gamma'\Delta'$
vertex is exactly extracted. 
This part  ${\widetilde {\cal H}}^{\mu}$ is connected with the internal
$\Delta$ radiation term ${\sc H}^{\mu}$ through 
current conservation as 
$k'_{\mu}A^{\mu}=k'_{\mu}{\widetilde {\cal H}}^{\mu}+k'_{\mu}{\sc H}^{\mu}=0$. 
This enables us to obtain a model independent relations 
between the $\Delta-\gamma'\Delta'$ vertex functions in  
${\sc H}^{\mu}$ and in ${\widetilde {\cal H}}^{\mu}$
and to determine the  magnetic dipole moment $\mu_{\Delta}$ of the $\Delta$  
via the anomalous magnetic moment of the proton $\mu_p$.

This paper consists of four sections and two appendixes. 
In  the next section 
 the current conservation conditions 
for the bremsstrahlung amplitude are derived
using the Ward-Takahashi identity.
In Sect. 3 an analytical extraction of the double $\Delta$ exchange
 diagram from the external particle radiation terms is given.
Using current conservation 
this term is combined with the internal $\Delta$ radiation diagram
(Fig.2B). This 
model independent relation between
the internal and external particle radiation terms
enables us to determine
 the  magnetic dipole momenta of $\Delta^+$ and $\Delta^{++}$  
through the anomalous magnetic moment of the proton.
The conclusions are presented in Sect. 4.
Appendix A  and appendix B give the formulas for projection
on the intermediate spin $3/2$ states and for the 
$\Delta-\gamma'\Delta'$ vertex functions.

\newpage

\begin{center}
                  {\bf 2. Ward-Takahashi identities for the 
pion-nucleon bremsstrahlung amplitude  }
\end{center}

\vspace{0.25cm}

\par
We consider the radiative pion-nucleon scattering 

$$\pi(p_{\pi})\ +\ N(p_{N})\Longrightarrow \gamma'(k')\ +\ \pi'(p'_{\pi})
\ +\ N'({p'_N})$$
with on mass shell momenta 
of the pi-meson ($p_{\pi}=(\sqrt{ {\bf p}_{\pi}^2+m_{\pi}^2},{\bf p}_{\pi})$,
${p'}_{\pi}=(\sqrt{ {\bf p'}_{\pi}^2+m_{\pi}^2},{\bf p'}_{\pi})$, nucleon
($p_{N}=(\sqrt{ {\bf p}_{N}^2+m_{N}^2},{\bf p}_{N})$,
${p'}_{N}=(\sqrt{ {\bf p'}_{N}^2+m_{N}^2},{\bf p'}_{N})$ and final photon
(${k'}^2=0$).
In the physically interesting case 
the energy-momentum of the final photon
is $k'_{\mu}=(p_N+p_{\pi}-p'_{\pi}-p'_N)_{\mu}$.

Following  the  derivation of the Ward-Takahashi 
identities (see e.g. ch. ${\bf 8.4.1}$ in
 the Itzykson and Zuber book\cite{IZ}) we start with 
the on shell amplitude 
$A_{\gamma'\pi' N'-\pi N}^{\mu}$

$${k'}_{\mu}A_{\gamma'\pi' N'-\pi N}^{\mu}
({\bf p'_{\pi},p'_N,k';p_{\pi},p_N})=
{\overline u}({\bf p'}_N)
(\gamma_{\nu} {p'_N}^{\nu}-m_N)({p'_{\pi}}^2-m_{\pi}^2)
{k'}_{\mu}{\tau}^{\mu}
(\gamma_{\nu} {p_N}^{\nu}-m_N)({p_{\pi}}^2-m_{\pi}^2)
 u({\bf p}_N),\eqno(2.1)$$
where the Green function ${\tau}^{\mu}$ is expressed 
via the photon source operator ${\cal J}^{\mu}(z)$ and 
the pion and the nucleon 
field operators $\Phi(x)$ and $\Psi(y)$ as

$$
{k'}_{\mu}{\tau}^{\mu}=
i\int d^4zd^4y'd^4x'd^4yd^4xe^{ik'z+ip'_{\pi}x'+ip'_Ny'-ip_{\pi}x-ip_Ny}
{{\partial}\over{\partial z^{\mu}} }
<0|{\sf T}\biggl(\Psi(y')\Phi(x'){\cal J}^{\mu}(z)
{\overline \Psi}(y)\Phi^+(x)\biggr)|0>.\eqno(2.2a)$$

Next we will use the well known relation for the time-ordered product 
of the quantum field operators

$${{\partial}\over{\partial z^{\mu}} }
<0|{\sf T}\biggl(\Psi(y')\Phi(x'){\cal J}^{\mu}(z)
{\overline \Psi}(y)\Phi^+(x)\biggr)|0>=
<0|{\sf T}\biggl(\Psi(y')\Phi(x')
\Bigl({{\partial}\over{\partial z^{\mu}} }{\cal J}^{\mu}(z)\Bigr)
{\overline \Psi}(y)\Phi^+(x)\biggr)|0>$$
$$+\delta(z_o-x'_o)<0|{\sf T}\biggl(\Psi(y')
\biggl[{\cal J}^{o}(z),\Phi(x')\biggr]
{\overline \Psi}(y)\Phi^+(x)\biggr)|0>$$
$$+\delta(z_o-y'_o)<0|{\sf T}\biggl(\Phi(x')
\biggl\{{\cal J}^{o}(z),\Psi(y')\biggr\}
{\overline \Psi}(y)\Phi^+(x)\biggr)|0>$$.
$$+\delta(z_o-x_o)<0|{\sf T}\biggl(\Psi(y')\Phi(x')
\biggl[{\cal J}^{o}(z),\Phi^+(x)\biggr]
{\overline \Psi}(y)\biggr)|0>$$
$$+\delta(z_o-y_o)<0|{\sf T}\biggl((\Psi(y')\Phi(x')
\biggl\{{\cal J}^{o}(z),{\overline \Psi}(y)\biggr\}
\Phi^+(x)\biggr)|0>$$.

and employ the equal-time commutation conditions  

$$\biggl\{ {\cal J}^{o}(z),\Psi(y')
\biggr\}\delta(z_o-y'_o)=-e_{N'}\delta^{(4)}(z-y')\Psi(y');\ \ \
\biggl\{ {\cal J}^{o}(z),{\overline \Psi}(y)
\biggr\}\delta(z_o-y_o)=e_N\delta^{(4)}(z-y){\overline \Psi}(y)\eqno(2.3a)$$

$$\biggl[ {\cal J}^{o}(z),\Phi(x')
\biggr]\delta(z_o-x'_o)=-e_{\pi'}\delta^{(4)}(z-x')\Phi(x');\ \ \
\biggl[ {\cal J}^{o}(z),{ \Phi^+}(x)
\biggr]\delta(z_o-x_o)=e_{\pi}\delta^{(4)}(z-x){ \Phi^+}(x),\eqno(2.3b)$$
where $e_{N'}$, $e_{\pi'}$, $e_N$ and  $e_{\pi}$ stand for the charge of the 
of the nucleons and pions in the final and initial states. In particular
$e_N=1,0$ for proton and neutron, and $e_{\pi}=\pm 1,0$ for pi-mesons.

As result of these relations, eq.(2.2a) for the Green function 
takes a form

$${k'}_{\mu}{\tau}^{\mu}=
-i\int d^4y'd^4x'd^4yd^4xe^{ip'_{\pi}x'+ip'_Ny'-ip_{\pi}+ip_Ny}
\biggl(e_{N'}e^{ik'y'}+e_{\pi'}e^{ik'x'}-
e_{N}e^{ik'y}-e_{\pi}e^{ik'x} \biggr)$$
$$<0|T\biggl(\Psi(y')\Phi(x')
{\overline \Psi}(y)\Phi^+(x)\biggr)|0>.\eqno(2.2b)$$

Equal-time commutators (2.3a,b) follow from the commutation relations
 between the electric charge operator $Q=\int d^3x {\cal J}^{o}(x)$ and the 
particle field operators with the charge $e$. 
These conditions 
express electric charge conservation for the local fields
i.e. they represent one of the first principles in the quantum field theory. 

Substituting expression (2.2b) into (2.1) we get

$${k'}_{\mu}A_{\gamma'\pi' N'-\pi N}^{\mu}
({\bf p'_{\pi},p'_N,k';p_{\pi},p_N})=-i
(2\pi)^4\ \delta^{(4)}(p'_N+p'_{\pi}+k'-p_{\pi}-p_N)$$
$$\Biggl[{\overline u}({\bf p'}_N)(\gamma_{\nu} {p'_N}^{\nu}-m_N)
{ {e_{N'}}\over{\gamma_{\nu} (p'_N+k')^{\nu}-m_N  }}
<out;{\bf p'}_{\pi}|J(0)|{\bf p}_{\pi}{\bf p}_N;in>$$
$$+( {p'}_{\pi}^2-m_{\pi}^2){{e_{\pi'}}\over{(p'_{\pi}+k')^2-m_{\pi}^2} }
<out;{\bf p'}_{N}|j_{\pi'}(0)|{\bf p}_{\pi}{\bf p}_N;in>$$
$$-
<out;{\bf p'}_{\pi}{\bf p'}_{N}|{\overline J}(0)|{\bf p}_{\pi};in>
{{e_{N}}\over{\gamma_{\nu} (p_N-k')^{\nu}-m_N} }(\gamma_{\nu} {p_N}^{\nu}-m_N)
u({\bf p}_N)$$
$$-<out;{\bf p'}_{\pi}{\bf p'}_{N}|j_{\pi}(0)|{\bf p}_{N};in>
{{e_{\pi}}\over{(p_{\pi}-k')^2-m_{\pi}^2}}
(p_{\pi}^2-m_{\pi}^2)
\Biggr]\eqno(2.4)$$
where $J(x)=(i\gamma_{\nu}\partial/\partial x_{\nu}-m_N)\Psi(x)$
and 
$j_{\pi}(x)=(\partial^2/\partial x^{\nu}\partial x_{\nu}-m_{\pi}^2)\Phi(x)$
denote the source operator of nucleon and pion.

For the on-mass shell external particles
eq.(2.4) vanishes. In particular, for $k'=0$ this expression  disappears 
due to cancellation of the on shell $\pi N$ amplitudes
in (2.4). Thus expression (2.4) presents
the current conservation 
condition for the on-mass shell bremsstrahlung amplitude

$${k'}_{\mu}
\Biggl[ A_{\gamma'\pi' N'-\pi N}^{\mu}
({\bf p'_{\pi},p'_N,k';p_{\pi},p_N})
\Biggr]_{on\ mass\ shell\ \pi',\ N',\ \pi,\ N}=0.\eqno(2.5a)$$

In  the off mass shell region 
for the four-momenta of the external 
pions and nucleons, where  
${q'_{\pi}}^2\ne m_{\pi}^2$ or ${q'_N}^2\ne m_{N}^2$ or
 $q_{\pi}^2\ne m_{\pi}^2$ or  $q_N^2\ne m_{N}^2$, 
eq.(2.4) is also  valid, but the current conservation condition is 
violated
$$\Biggl[{k'}_{\mu}A_{\gamma'\pi' N'-\pi N}^{\mu}
( q'_{\pi},q'_N;q_{\pi},q_N)\Biggr]_{off\ mass\ shell\ \pi',\ N',\ \pi,\ N}
\ne 0.\eqno(2.5b)$$

It is convenient to extract the full energy-momentum conservation 
$\delta$ function from the  radiative $\pi N$ scattering 
amplitude $A_{\gamma'\pi' N'-\pi N}^{\mu}$ and introduce the corresponding
nonsingular amplitude
$<out;{\bf p'}_{N}{\bf p'}_{\pi}|J^{\mu}(0)|{\bf p}_{\pi}{\bf p}_N;in>$

$${k'}_{\mu}A_{\gamma'\pi' N'-\pi N}^{\mu}=-i
(2\pi)^4\ \delta^{(4)}(p'_N+p'_{\pi}+k'-p_{\pi}-p_N)
{k'}_{\mu}
<out;{\bf p'}_{N}{\bf p'}_{\pi}|J^{\mu}(0)|{\bf p}_{\pi}{\bf p}_N;in>,
\eqno(2.6)$$

Afterwards using the identity $a/(a+b)\equiv 1- b/(a+b)$
in eq. (2.4) we obtain

$${k'}_{\mu}
<out;{\bf p'}_{N}{\bf p}_{\pi'}|J^{\mu}(0)|{\bf p}_{\pi}{\bf p}_N;in>=
\Biggl[{\cal B}_{\gamma'\pi' N'-\pi N}+
{k'}_{\mu}{\cal E}_{\gamma'\pi' N'-\pi N}^{\mu}
\Biggr]_{on\ mass\ shell\ \pi',\ N',\ \pi,\ N}=0,\eqno(2.7a)$$

$${k'}_{\mu}
<{ q'}_{N}{ q'}_{\pi}|J^{\mu}(0)|{ q}_{\pi}{q}_N>=
\Biggl[{\cal B}_{\gamma'\pi' N'-\pi N}+
{k'}_{\mu}{\cal E}_{\gamma'\pi' N'-\pi N}^{\mu}\Biggr|_{off\ mass\ shell
\  \pi',\ or\ N',\ or\ \pi,or\ N}\ne 0,\eqno(2.7b)$$

where

$${\cal B}_{\gamma'\pi' N'-\pi N}=
e_{N'}{\overline u}({\bf p'}_N)
<out;{\bf p'}_{\pi}|J(0)|{\bf p}_{\pi}{\bf p}_N;in>
+e_{\pi'}<out;{\bf p'}_{N}|j_{\pi'}(0)|{\bf p}_{\pi}{\bf p}_N;in>$$
$$-e_N<out;{\bf p'}_{\pi}{\bf p'}_{N}|{\overline J}(0)|{\bf p}_{\pi};in>
u({\bf p}_N)-e_{\pi}<out;{\bf p'}_{\pi}{\bf p'}_{N}|j_{\pi}(0)|{\bf p}_{N};in>,
\eqno(2.8a)$$

$${\cal E}_{\gamma'\pi' N'-\pi N}^{\mu}=
-\Biggl[{\overline u}({\bf p'}_N)\gamma^{\mu}
{{\gamma_{\nu} (p'_N+k')^{\nu}+m_N }\over{
2p'_Nk'}}e_{N'}
<out;{\bf p'}_{\pi}|J(0)|{\bf p}_{\pi}{\bf p}_N;in>$$
$$+(2{p'}_{\pi}+k')^{\mu}{{e_{\pi'}}
\over{ {2p'_{\pi}k' }} }
<out;{\bf p'}_{N}|j_{\pi'}(0)|{\bf p}_{\pi}{\bf p}_N;in>$$
$$-e_N
<out;{\bf p'}_{\pi}{\bf p'}_{N}|{\overline J}(0)|{\bf p}_{\pi};in>
{ {\gamma_{\nu} (p_N-k')^{\nu}+m_N}
\over{2p_Nk'} }\gamma^{\mu}
u({\bf p}_N)$$
$$-<out;{\bf p'}_{\pi}{\bf p'}_{N}|j_{\pi}(0)|{\bf p}_{N};in>
{{e_{\pi}}\over{2p_{\pi}k'} }
(2p_{\pi}-k')^{\mu}\Biggr]\eqno(2.8b)$$

 Amplitude  ${\cal E}_{\gamma'\pi' N'-\pi N}^{\mu}$  describes the 
$\pi N\to\gamma'\pi'N'$ reaction with  photon emission
through the $N-\gamma' N'$ and $\pi-\gamma'\pi$ vertex functions
in the tree approximation.  ${\cal E}_{\gamma'\pi' N'-\pi N}^{\mu}$ (2.8b)
can be described as a combination of the Feynman diagrams  
with external particle radiation as depicted on Fig.1. 
This external particle radiation amplitude is the starting formula
for the derivation of the  soft photon  theorem \cite{Low}.
Various applications of this method are given in 
\cite{Adler,Mink,Liou,Ding,Lin}.
Diagrams of Fig. 1 are responsible for the
infrared  behavior of the bremsstrahlung amplitude. 
The soft photon theorem enables us to restrict  calculations 
of bremsstrahlung reactions to the leading 
diagrams on Fig. 1 and Fig. 2B which insures the validity of current 
conservation. 

\vspace{0.7cm}

\begin{figure}[htb]
\includegraphics[width=14.0cm]{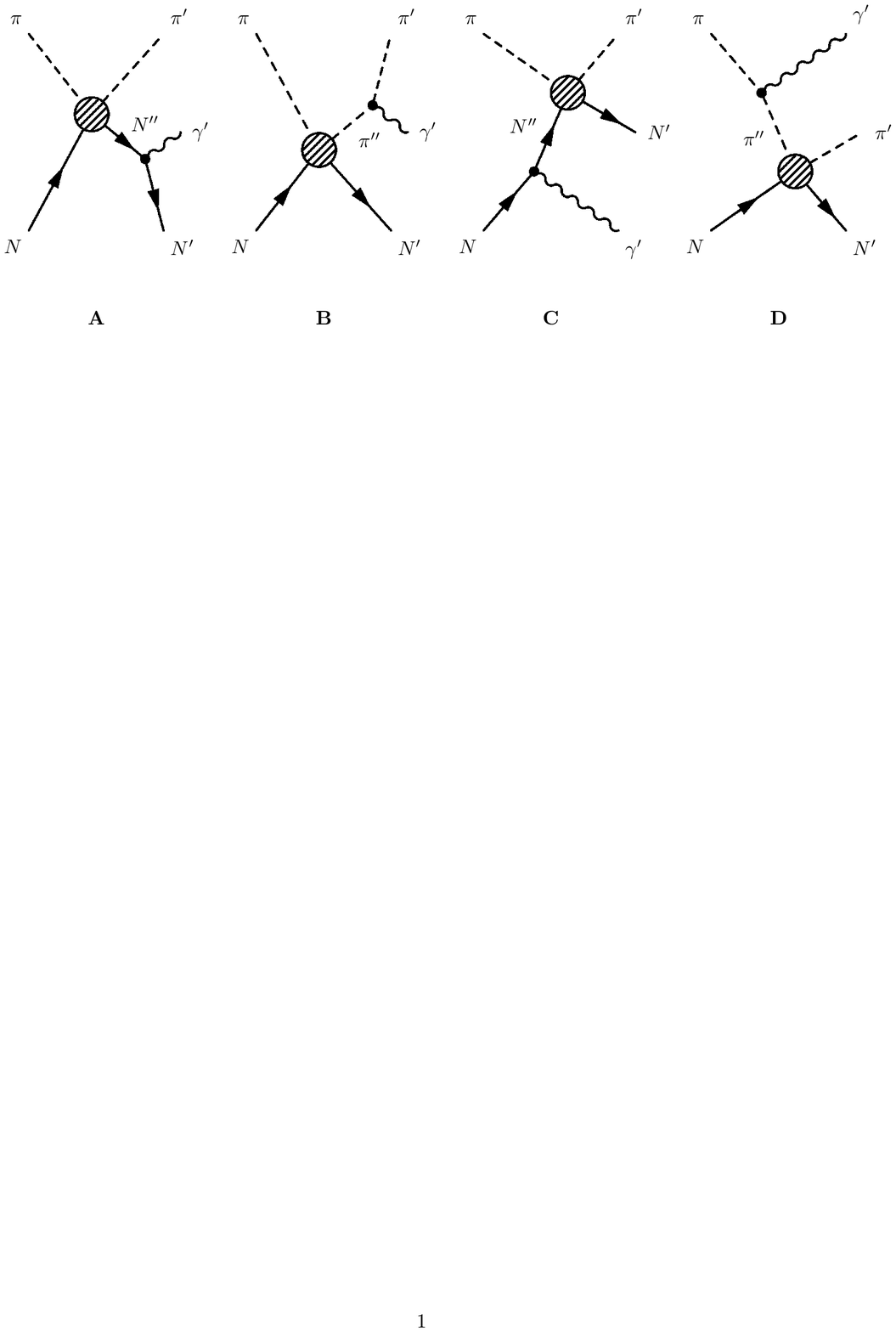}
\vspace{-15.0cm} 
\caption{{\protect\footnotesize {\it
Diagrams describing $\pi N$ bremsstrahlung 
with photon emission by the external 
nucleons (A,C) and by the external pions (B,D) in
${\cal E}_{\gamma'\pi' N'-\pi N}^{\mu}$ (2.8b).
The dashed circle indicates the off shell 
$\pi N$ elastic scattering amplitudes (2.9a,b,c,d).
}} }
\label{fig:one}
\end{figure}

\vspace{5mm}

\vspace{0.5cm}

\begin{figure}[htb]
\includegraphics[width=10.0cm]{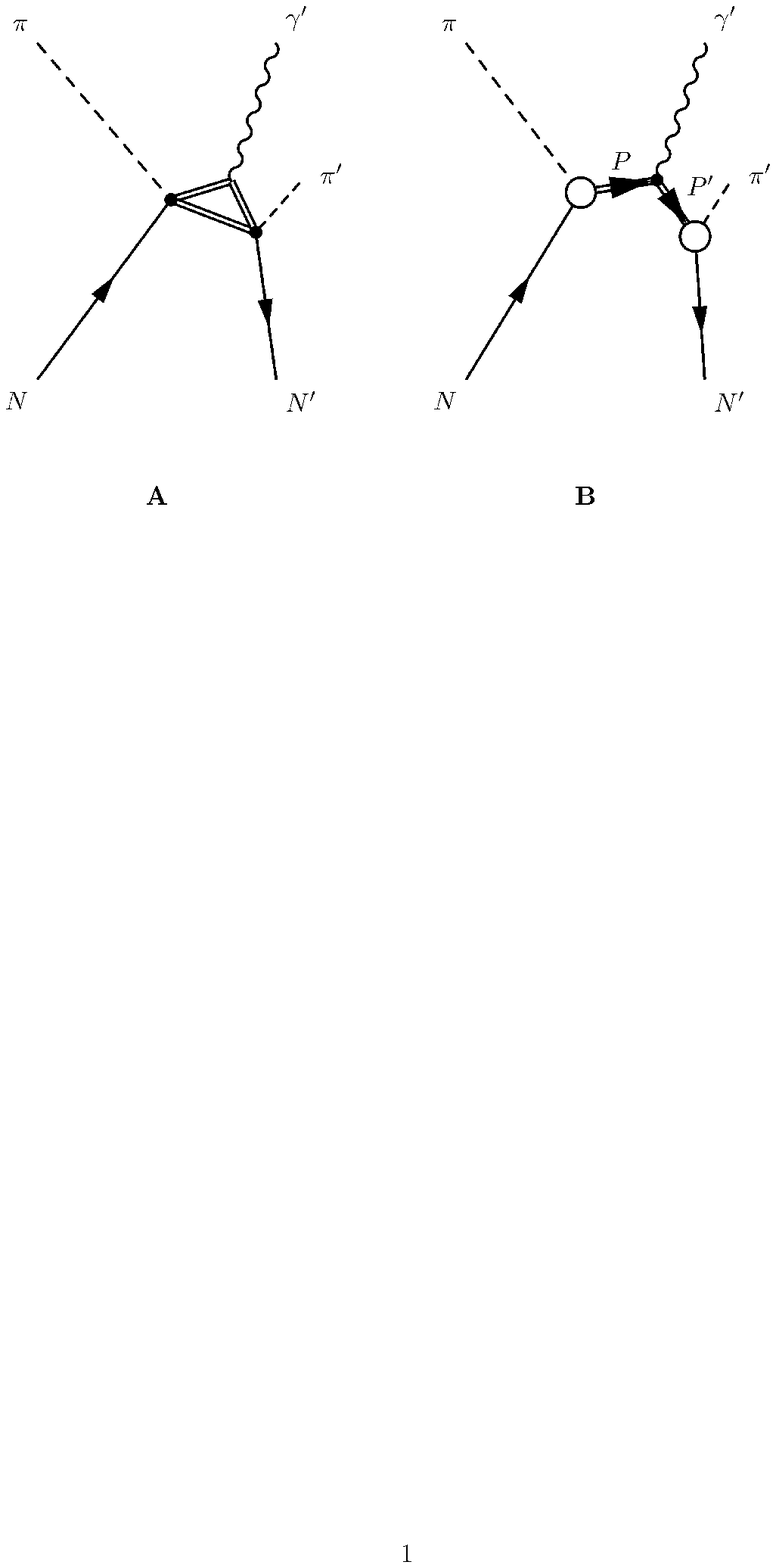}
\vspace{-12.0cm} 
\caption{{\protect\footnotesize {\it
 Diagram A presents  a symbolic description of the internal particle  
radiation amplitude ${\cal I}_{\gamma'\pi' N'-\pi N}^{\mu}$
in eq.(3.17a). Diagram  B describes
the photon emission from the intermediate $\pi N$ cluster 
with the four momentums $P$ and $P'$.}}}
\label{fig:two}
\end{figure}


The present derivation of current conservation  
(2.7a,b) is based on  the Ward-Takahashi identity (2.2b)
and is not restricted
by the  energy $k'=|{\bf k'}_{\gamma}|$ of the final photon.
An interesting property is that the expressions (2.4) and (2.8b) 
contain the electromagnetic 
form factors of the external particles in the tree approximation.
This is a result of the equal time commutation rules (2.3a,b) which  
follow from charge conservation. 
Thus current conservation 
(2.7a) is valid only for the on mass shell external particles and this
condition is fulfilled for an arbitrary energy of the initial and final 
particles. 
But in the off mass shell region
  we get instead of (2.7a) the similar condition (2.7b) .

The $\pi N$ scattering amplitudes in (2.8a,b) 
are functions of the three on-mass shell momenta 
from which one can construct only three  independent 
Lorentz-invariant (Mandelstam) variables.
Therefore we have

$${\overline u}({\bf p'}_N)
<out;{\bf p'}_{\pi}|J(0)|{\bf p}_{\pi}{\bf p}_N;in>
={\cal T}_{N'}\biggl( (p'_{\pi}-p_{\pi}-p_N)^2;s,t_{\pi}\biggr)=
{\cal T}_{N'}\biggl( m_N^2+2k'p'_N;s,t_{\pi}\biggr)
\eqno(2.9a)$$

$$<out;{\bf p'}_{N}|j_{\pi'}(0)|{\bf p}_{\pi}{\bf p}_N;in>
={\cal T}_{\pi'}\biggl( (p'_{N}-p_{\pi}-p_N)^2;s,t_{N}\biggr)=
{\cal T}_{\pi'}\biggl( m_{\pi}^2+2k'p'_{\pi};s,t_{N}\biggr)\eqno(2.9b)$$

$$<out;{\bf p'}_{\pi}{\bf p'}_{N}|{\overline J}(0)|{\bf p}_{\pi};in>
u({\bf p}_N)
={\cal T}_{N}\biggl(s',t_{\pi};(p'_{\pi}+p'_N-p_{\pi})^2\biggr)=
{\cal T}_{N}\biggl(s',t_{\pi}; m_N^2-2k'p_N\biggr)\eqno(2.9c)$$

$$<out;{\bf p'}_{\pi}{\bf p'}_{N}|j_{\pi}(0)|{\bf p}_{N};in>
={\cal T}_{\pi}\biggl(s',t_N;(p'_{\pi}+p'_N-p_{\pi})^2\biggr)=
{\cal T}_{\pi}\biggl(s',t_N; m_{\pi}^2-2k'p_{\pi}\biggr).\eqno(2.9d)$$

Amplitudes (2.9a,b,c,d) can be represented as a sum of the Feynman diagrams 
for the elastic $\pi N$ scattering reactions, where three external particles 
being on mass shell and the four momentum of the
fourth particle is determined via the energy-momentum 
conservation for the bremsstrahlung amplitude
(2.6). The related invariant variable are

$$s'=(p'_N+p'_{\pi})^2;\ \ \ s=(p_N+p_{\pi})^2=(p'_N+p'_{\pi}+k')^2
=s'+2k'(p'_N+p'_{\pi})=s'+2k'(p_N+p_{\pi})\eqno(2.10a)$$ 

$$t_N=(p'_N-p_N)^2;\ \ \ t_{\pi}=(p'_{\pi}-p_{\pi})^2\eqno(2.10b)$$

with the following relations between them 

$$t_{\pi}+(p'_{\pi}-p_N)^2+s
=m_{\pi}^2+2m_{N}^2+(p'_{\pi}-p_{\pi}-p_N)^2,\eqno(2.11a)$$

$$t_N+(p'_N-p_{\pi})^2+s=m_N^2+2m_{\pi}^2+(p'_N-p_{\pi}-p_N)^2,\eqno(2.11b)$$

$$t_{\pi}+(p_{\pi}-p'_N)^2+s'
=m_{\pi}^2+2m_{N}^2+(p_{\pi}-p'_{\pi}-p'_N)^2,\eqno(2.11c)$$

$$t_N+(p_N-p'_{\pi})^2+s'=m_N^2+2m_{\pi}^2+(p_N-p'_{\pi}-p'_N)^2.
\eqno(2.11d)$$

Next we represent amplitude (2.8b) in the time-ordered three-dimensional
form which contains one on mass shell nucleon (Fig.1A,C) and one on
mass shell pion (Fig.1B,D) exchange diagrams  

$${\widetilde {\cal E}}_{\gamma'\pi' N'-\pi N}^{\mu}=
-\Biggl[ {{
{\overline u}({\bf p'}_N)\Bigl[(2p'_N+k')^{\mu}
-i\sigma^{\mu\nu}k'_{\nu}\Bigr] }\over{ 2p'_Nk'} }
{{ (p'_N+k')_{\sigma}\gamma^{\sigma}+m_N}\over {2m_N}}
 e_{N'}<out;{\bf p'}_{\pi}|J(0)|{\bf p}_{\pi}{\bf p}_N;in>
$$
$$+(2{p'}_{\pi}+k')^{\mu}{ {e_{\pi'}}
\over{ {2p'_{\pi}k' }} }
<out;{\bf p'}_{N}|j_{\pi'}(0)|{\bf p}_{\pi}{\bf p}_N;in>
$$
$$
-e_N<out;{\bf p'}_{\pi}{\bf p'}_{N}|{\overline J}(0)|{\bf p}_{\pi};in>
{{ (p_N-k')_{\sigma}\gamma^{\sigma}+m_N}\over {2m_N}}
{ {\Bigl[(2p_N-k')^{\mu}-i\sigma^{\mu\nu}k'_{\nu}
\Bigr]u({\bf p}_N) }\over{2p'_Nk'} }$$
$$-<out;{\bf p'}_{\pi}{\bf p'}_{N}|j_{\pi}(0)|{\bf p}_{N};in>
{{e_{\pi}}\over{2p_{\pi}k'} }
(2p_{\pi}-k')^{\mu} 
\Biggr],\eqno(2.12)$$

where relations 
${\overline u}({\bf p'}_N)\gamma^{\mu} 
( \gamma_{\nu} {p'_N}^{\nu}+m_N)=
{\overline u}({\bf p'}_N)2{ p'_N}^{\mu}$ and
${\overline u}({\bf p'}_N)\gamma^{\mu} 
\gamma^{\nu} k'_{\nu}=
{k'}^{\mu}{\overline u}({\bf p'}_N)-i{\overline u}({\bf p'}_N)
\sigma^{\mu\nu}k'_{\nu}
\Bigl[ u({\bf p'_N+k'}){\overline u}({\bf p'_N+k'})
- v({\bf p'_N+k'}){\overline v}({\bf p'_N+k'})\Bigr]$
are used. 
For the sake of simplicity contributions of the 
antinucleon intermediate states in (2.12) are omitted. Therefore 
${\widetilde {\cal E}}_{\gamma'\pi' N'-\pi N}^{\mu}
\ne{\cal E}_{\gamma'\pi' N'-\pi N}^{\mu}$ and instead of (2.7a)
we get

$$\biggl[{k'}_{\mu}
<out;{\bf p'}_{N}{\bf p}_{\pi'}|J^{\mu}(0)|{\bf p}_{\pi}{\bf p}_N;in>=
{\widetilde {\cal B}}_{\gamma'\pi' N'-\pi N}+
{k'}_{\mu} {\widetilde {\cal E}}_{\gamma'\pi' N'-\pi N}^{\mu}
\biggr]^{without\ \pi-\pi N{\overline N}\ transition}
_{on\ mass\ shell\ \pi',\ N',\ \pi,\ N}=0,\eqno(2.13)$$

where

$${\widetilde {\cal B}}_{\gamma'\pi' N'-\pi N}
=e_{N'}{\overline u}({\bf p'}_N)
{{ (p'_N+k')_{\sigma}\gamma^{\sigma}+m_N}\over {2m_N}}
<out;{\bf p'}_{\pi}|J(0)|{\bf p}_{\pi}{\bf p}_N;in>
+e_{\pi'}<out;{\bf p'}_{N}|j_{\pi'}(0)|{\bf p}_{\pi}{\bf p}_N;in>$$
$$-e_N<out;{\bf p'}_{\pi}{\bf p'}_{N}|{\overline J}(0)|{\bf p}_{\pi};in>
{{ (p_N-k')_{\sigma}\gamma^{\sigma}+m_N}\over {2m_N}}
u({\bf p}_N)-e_{\pi}<out;{\bf p'}_{\pi}{\bf p'}_{N}|j_{\pi}(0)|{\bf p}_{N};in>,
\eqno(2.14)$$

Conditions (2.7a), (2.8a,b) and (2.12) establish
a relationships between the $\pi N$ brems\-strahlung amplitude and the
off mass shell elastic $\pi N$ scattering amplitudes.   
The gauge terms in (2.12)  are proportional to ${k'}^{\mu}$ and 
$\sigma^{\mu\nu}{k'}_{\nu}$. These terms  modify the 
Green function $\tau^{\mu}$ in (2.1). But for the on shell 
amplitude they can be ignored. 

The gauge terms in 
${\cal E}_{\gamma'\pi' N'-\pi N}^{\mu}$ (2.12) (Fig. 1)
must contain contribution from the anomalous magnetic moment of 
the nucleon.
For this aim one can use a gauge transformation  
$\gamma_{\mu}\Longrightarrow \gamma_{\mu}-i
\mu/2m_N\sigma_{\mu\nu}{k'}^{\nu}$ which replaces 
$i{\overline u}({\bf p'}_N)
\sigma^{\mu\nu}k'_{\nu}u({\bf p'}_N)$ via 
$i\mu_{N'}{\overline u}({\bf p'}_N)
\sigma^{\mu\nu}k'_{\nu}u({\bf p'}_N)$. Then we obtain 

$${\widetilde {\cal E}}_{\gamma'\pi' N'-\pi N}^{\mu}=
-\Biggl[ {{
{\overline u}({\bf p'}_N)\Bigl[(2p'_N+k')^{\mu}
-i\mu_{N'}\sigma^{\mu\nu}k'_{\nu}\Bigr] }\over{ 2p'_Nk'} }
{{ (p'_N+k')_{\sigma}\gamma^{\sigma}+m_N}\over {2m_N}}
 e_{N'}<out;{\bf p'}_{\pi}|J(0)|{\bf p}_{\pi}{\bf p}_N;in>
$$
$$+(2{p'}_{\pi}+k')^{\mu}{ {e_{\pi'}}
\over{ {2p'_{\pi}k' }} }
<out;{\bf p'}_{N}|j_{\pi'}(0)|{\bf p}_{\pi}{\bf p}_N;in>
$$
$$
-e_N<out;{\bf p'}_{\pi}{\bf p'}_{N}|{\overline J}(0)|{\bf p}_{\pi};in>
{{ (p_N-k')_{\sigma}\gamma^{\sigma}+m_N}\over {2m_N}}
{ {\Bigl[(2p_N-k')^{\mu}-i\mu_{N}\sigma^{\mu\nu}k'_{\nu}
\Bigr]u({\bf p}_N) }\over{2p'_Nk'} }$$
$$-<out;{\bf p'}_{\pi}{\bf p'}_{N}|j_{\pi}(0)|{\bf p}_{N};in>
{{e_{\pi}}\over{2p_{\pi}k'} }
(2p_{\pi}-k')^{\mu} 
\Biggr],\eqno(2.15)$$

 ${\widetilde {\cal E}}_{\gamma'\pi' N'-\pi N}^{\mu}$
 (2.15) describes the  radiative $\pi N$ scattering amplitude
with the external particle radiation, 
where the magnetic momenta of external nucleons determine the form of 
the gauge terms at $\sigma^{\mu\nu}k'_{\nu}$.

\vspace{0.25cm}

\begin{center}
                  {\bf 3. Internal and external  particle radiation parts}
\end{center}

\vspace{0.25cm}

It is convenient to introduce the total and relative momenta
$$P=p_{\pi}+p_N;\ \ \ 
p={{\alpha_{\pi}p_N-\alpha_{N}p_{\pi}}\over{\alpha_{\pi}+\alpha_N}};\ \ \  
p_N={{\alpha_{N}P}\over{\alpha_{\pi}+\alpha_N}}+p,\ \ \
p_{\pi}={{\alpha_{\pi}P}\over{\alpha_{\pi}+\alpha_N}}-p,\eqno(3.1a)$$

$$P'=p'_{\pi}+p'_N;\ \ \ 
p'={{\alpha'_{\pi}p'_N-\alpha'_{N}p'_{\pi}}
\over{\alpha'_{\pi}+\alpha'_N}};\ \ \  
p'_N={{\alpha'_{N}P'}\over{\alpha'_{\pi}+\alpha'_N}}+p',\ \ \
p'_{\pi}={{\alpha'_{\pi}P'}\over{\alpha'_{\pi}+\alpha'_N}}-p',\eqno(3.1b)$$

where

$$\alpha_{N}=k'_{\nu}p_{N}^{\nu},\ \ \
\alpha_{\pi}=k'_{\nu}p_{\pi}^{\nu};\ \ \
\alpha'_{N}=k'_{\nu}{p'_{N}}^{\nu},\ \ \
\alpha'_{\pi}=k'_{\nu}{p'_{\pi}}^{\nu}.\eqno(3.1c)$$

The relative moments $p$ and $p'$ are transverse to $k'$

$$k'_{\nu}{p}^{\nu}=0;\ \ \ k'_{\nu}{p'}^{\nu}=0.\eqno(3.2)$$

Therefore we can  extract from  
${\widetilde {\cal E}}_{\gamma'\pi' N'-\pi N}^{\mu}$  (2.15) 
the transverse part ${\cal C}_{\gamma'\pi' N'-\pi N}^{\mu}$

$${\widetilde {\cal E}}_{\gamma'\pi' N'-\pi N}^{\mu}=
{\cal M}
_{\gamma'\pi' N'-\pi N}^{\mu}+{\cal C}_{\gamma'\pi' N'-\pi N}^{\mu}
\eqno(3.3a)$$

and

$$
{k'}_{\mu}{\cal M}_{\gamma'\pi' N'-\pi N}^{\mu}
=-{\widetilde {\cal B}}_{\gamma'\pi' N'-\pi N};\ \ \ 
{k'}_{\mu}{\widetilde{\cal E}}_{\gamma'\pi' N'-\pi N}^{\mu}
=-{\widetilde {\cal B}}_{\gamma'\pi' N'-\pi N}\ \ \ and\ \ \  
{k'}_{\mu}{\cal C}_{\gamma'\pi' N'-\pi N}^{\mu}
=0. \eqno(3.3b)$$

Using (2.15) and (3.1a,b,c) we get

$${\cal M}_{\gamma'\pi' N'-\pi N}^{\mu}=
-{1\over{s-s'}}\Biggl[ 
{\overline u}({\bf p'_N})\Bigl[2e_{N'}{P'}^{\mu}
-i\mu_{N'}\sigma^{\mu\nu}k'_{\nu}\Bigr]u({\bf p'_N+k'})
{\overline u}({\bf p'_N+k'})<out;{\bf p'}_{\pi}|J(0)|{\bf p}_{\pi}{\bf p}_N;in>
$$
$$-<out;{\bf p'}_{\pi}{\bf p'}_{N}|{\overline J}(0)|{\bf p}_{\pi};in>
u({\bf p_N-k'}){\overline u}({\bf p_N-k'})\Bigl[2e_N(P)^{\mu}
-i\mu_N\sigma^{\mu\nu}k'_{\nu}
\Bigr]u({\bf p}_N)\Biggr]$$
$$-{1\over{s-s'}}\Biggl[2e_{\pi'}{P'}^{\mu}
<out;{\bf p'}_{N}|j_{\pi'}(0)|{\bf p}_{\pi}{\bf p}_N;in>
-2e_{\pi}P^{\mu}
<out;{\bf p'}_{\pi}{\bf p'}_{N}|j_{\pi}(0)|{\bf p}_{N};in>
\Biggr]\eqno(3.4a)$$

$${\cal C}_{\gamma'\pi' N'-\pi N}^{\mu}=
-\Biggl[ e_{N'}{{{p'}^{\mu}}\over{\alpha'_N}}
{\overline u}({\bf p'}_N)u({\bf p'_N+k'})
{\overline u}({\bf p'_N+k'})<out;{\bf p'}_{\pi}|J(0)|{\bf p}_{\pi}{\bf p}_N;in>
$$
$$-e_{\pi'}{{{p'}^{\mu}}\over{\alpha'_{\pi} }}
<out;{\bf p'}_{N}|j_{\pi'}(0)|{\bf p}_{\pi}{\bf p}_N;in>
+e_{\pi}{{{p}^{\mu}}\over{\alpha_{\pi} } }
<out;{\bf p'}_{\pi}{\bf p'}_{N}|j_{\pi}(0)|{\bf p}_{N};in>$$
$$-e_{N}{{{p}^{\mu}}\over{\alpha_N}}
<out;{\bf p'}_{\pi}{\bf p'}_{N}|{\overline J}(0)|{\bf p}_{\pi};in>
u({\bf p_N-k'}){\overline u}({\bf p_N-k'})u({\bf p}_N)
\Biggr]\eqno(3.4b)$$

One can separate an other gauge term in (3.4b)
 introducing a new total four-momenta $P_{\pm}$
instead of $P$ and $P'$ 

$$P_{\pm}={1\over2}(P\pm P'),\ \ where \ P=P_+ +P_-;\ \ \ P'=P_+-P_-
\ \ \ and\ \ \ {P_-}^{\mu}={1\over 2}{k'}^{\mu}.\eqno(3.5a)$$

This enables us to pick out 
another gauge term ${\cal F}_{\gamma'\pi' N'-\pi N}^{\mu}$

$${\cal M}_{\gamma'\pi' N'-\pi N}^{\mu}=
  {\cal D}_{\gamma'\pi' N'-\pi N}^{\mu}+
  {\cal F}_{\gamma'\pi' N'-\pi N}^{\mu},\eqno(3.6)$$

where

$${\cal D}_{\gamma'\pi' N'-\pi N}^{\mu}=
-{1\over{s-s'}}
\Biggl[ {\overline u}({\bf p'_N})\Bigl[e_{N'}(P+P')^{\mu}
-i\mu_{N'}\sigma^{\mu\nu}k'_{\nu}\Bigr]u({\bf p'_N+k'}) 
{\overline u}({\bf p'_N+k'})<out;{\bf p'}_{\pi}|J(0)|{\bf p}_{\pi}{\bf p}_N;in>
$$
$$-<out;{\bf p'}_{\pi}{\bf p'}_{N}|{\overline J}(0)|{\bf p}_{\pi};in>
u({\bf p_N-k'}){\overline u}({\bf p_N-k'})\Bigl[e_N(P'+P)^{\mu}
-i\mu_N\sigma^{\mu\nu}k'_{\nu}
\Bigr]u({\bf p}_N)\Biggr]$$
$$-{1\over{s-s'}}\Biggl[e_{\pi'}{(P+P')}^{\mu}
<out;{\bf p'}_{N}|j_{\pi'}(0)|{\bf p}_{\pi}{\bf p}_N;in>
-e_{\pi}{(P+P')}^{\mu}
<out;{\bf p'}_{\pi}{\bf p'}_{N}|j_{\pi}(0)|{\bf p}_{N};in>
\Biggr],\eqno(3.7a)$$

$${\cal F}_{\gamma'\pi' N'-\pi N}^{\mu}={{{ k'}^{\mu}}\over{s-s'}}\Biggl[ 
 {\overline u}({\bf p'}_N)
{{ (p'_N+k')_{\sigma}\gamma^{\sigma}+m_N}\over {2m_N}}
e_{N'}<out;{\bf p'}_{\pi}|J(0)|{\bf p}_{\pi}{\bf p}_N;in>$$
$$+e_{\pi'}<out;{\bf p'}_{N}|j_{\pi'}(0)|{\bf p}_{\pi}{\bf p}_N;in>$$
$$+e_N<out;{\bf p'}_{\pi}{\bf p'}_{N}|{\overline J}(0)|{\bf p}_{\pi};in>
{{ (p_N-k')_{\sigma}\gamma^{\sigma}+m_N}\over {2m_N}}
u({\bf p}_N)+e_{\pi}
<out;{\bf p'}_{\pi}{\bf p'}_{N}|j_{\pi}(0)|{\bf p}_{N};in>
\Biggr],\eqno(3.7b)$$

A projection on the spin $3/2$ intermediate states in 
the $\gamma N-N$ vertex function, as shown in appendix A,
generates the following redefinition of the expression (3.7a)
${\cal D}_{\gamma'\pi' N'-\pi N}^{\mu}\Longrightarrow
{\cal H}_{\gamma'\pi' N'-\pi N}^{\mu}$

$${\cal H}_{\gamma'\pi' N'-\pi N}^{\mu}=
{{ (p'_N)_{a} (p_N)_d {(P+P')}^{\mu}}\over{ 
 (p'_N{\bf .}p_N)(s-s')}}
{\overline u}({\bf p'_N})i\gamma_5u^a({\bf P'_{\Delta}})\Bigl\{
{\overline u}^b({\bf P'_{\Delta}})g_{bc}u^c({\bf P_{\Delta}})\Bigr\}
{\overline u}^d({\bf P_{\Delta}})i\gamma_5u({\bf p_N})$$
$$\Biggl[{\overline u}({\bf p_N})
{{ (p'_N+k')_{\sigma}\gamma^{\sigma}+m_N}\over {2m_N}}
e_{N'}<out;{\bf p'}_{\pi}|J(0)|{\bf p}_{\pi}{\bf p}_N;in>$$
$$-e_{N}<out;{\bf p'}_{\pi}{\bf p'}_{N}|{\overline J}(0)|{\bf p}_{\pi};in>
{{ (p_N-k')_{\sigma}\gamma^{\sigma}+m_N}\over {2m_N}}u({\bf p'_N})$$
$$
+{\overline u}({\bf p_N})u({\bf p'_N})
e_{\pi'}<out;{\bf p'}_{N}|j_{\pi'}(0)|{\bf p}_{\pi}{\bf p}_N;in>
-e_{\pi}<out;{\bf p'}_{\pi}{\bf p'}_{N}|j_{\pi}(0)|{\bf p}_{N};in>
{\overline u}({\bf p_N})u({\bf p'_N})
\Biggr]$$
$$+{{ (p'_N)_{a} (p_N)_d }\over{ 
 (p'_N{\bf .}p_N)(s-s')}}
{\overline u}({\bf p'_N})i\gamma_5u^a({\bf P'_{\Delta}})\Bigl\{
{\overline u}^b({\bf P'_{\Delta}})g_{bc}
(-i\sigma^{\mu\nu}k'_{\nu})u^c({\bf P_{\Delta}})\Bigr\}
{\overline u}^d({\bf P_{\Delta}})i\gamma_5u({\bf p_N})$$
$$\Biggl[ {\overline u}({\bf p_N})
{{ (p'_N+k')_{\sigma}\gamma^{\sigma}+m_N}\over {2m_N}}
\mu_{N'}<out;{\bf p'}_{\pi}|J(0)|{\bf p}_{\pi}{\bf p}_N;in>$$
$$-\mu_{N}<out;{\bf p'}_{\pi}{\bf p'}_{N}|{\overline J}(0)|{\bf p}_{\pi};in>
{{ (p_N-k')_{\sigma}\gamma^{\sigma}+m_N}\over {2m_N}}u({\bf p'_N})\Biggr]
,\eqno(3.8)$$

where $(p'_N{\bf .}p_N)=(p'_{N})_{\sigma} (p_N)^{\sigma}$ and
in $u^b({\bf P_{\Delta}})$ $u^c({\bf P'_{\Delta}})$
 the four-momentums $P_{\Delta}$ and $P'_{\Delta}$ with complex mass 
$\Sigma(s)=M_{\Delta}(s)-{{i\Gamma_{\Delta}(s)}\over 2}$
are assumed   

$$P^o_{\Delta}(s)=\sqrt{
(M_{\Delta}(s)-{{i\Gamma_{\Delta}(s)}\over 2})^2
+{\bf P}_{\Delta}^2};\ \ \ \ \ \ \
\ {\bf P}_{\Delta}={\bf p}_N+{\bf p}_{\pi}\eqno(3.9a)$$

$${P'}^o_{\Delta}(s')=\sqrt{(M_{\Delta}(s')-{{i\Gamma_{\Delta}(s')}\over 2})^2
+{\bf P'}_{\Delta}^2 };\ \ \ \ \ \ 
{\bf P'}_{\Delta}={\bf p'}_N+{\bf p'}_{\pi}.\eqno(3.9b)$$

${\cal H}_{\gamma'\pi' N'-\pi N}^{\mu}$ has the $s$-channel two-body poles at
$s=s'$ and the $\Delta$ resonance poles
in the complex energy region from the $\pi N$ amplitudes. 
Therefore ${\cal H}_{\gamma'\pi' N'-\pi N}^{\mu}$
can be represented in the form of the  double $\Delta$ (or $\pi N$ cluster)  
exchange term (Fig. 2B) if we use the
$s$-channel pole approximation for the $\pi N$ amplitudes (2.9a,b,c,d)
in the $\Delta$ resonance region

$$
{\overline u}({\bf p_N}){{ (p'_N+k')_{\sigma}\gamma^{\sigma}+m_N}\over {2m_N}}
{\overline u}({\bf p'}_N)
<out;{\bf p'}_{\pi}|J(0)|{\bf p}_{\pi}{\bf p}_N;in>
\Longrightarrow
{{ {\sc r}_{N'}}\over{
{p}_{\pi}^{o}+{p}_{N}^{o}-{P'}^o_{\Delta}(s)}}
;\eqno(3.10a)$$

$$
{\overline u}({\bf p_N})u({\bf p'_N})
<out;{\bf p'}_{N}|j_{\pi'}(0)|{\bf p}_{\pi}{\bf p}_N;in>\Longrightarrow
{{ {\sc r}_{\pi'}}\over{
{p}_{\pi}^{o}+{p}_{N}^{o}-
{P}^o_{\Delta}(s) }};\eqno(3.10b)$$

$$<out;{\bf p'}_{\pi}{\bf p'}_{N}|{\overline J}(0)|{\bf p}_{\pi};in>
u({\bf p}_N){{ (p_N-k')_{\sigma}\gamma^{\sigma}+m_N}\over {2m_N}}
u({\bf p'_N})\Longrightarrow
{{ {\sc r}_{N} }\over{
{p'}_{\pi}^{o}+{p'_{N}}^{o}-
{P'}^o_{\Delta}(s')} };
\eqno(3.10c)$$

$$
<out;{\bf p'}_{\pi}{\bf p'}_{N}|j_{\pi}(0)|{\bf p}_{N};in>
{\overline u}({\bf p_N})u({\bf p'_N})\Longrightarrow
{{ {\sc r}_{\pi} }\over{
{p'}_{\pi}^{o}+{p'}_{N}^{o}-{P'}^o_{\Delta}(s') }},
\eqno(3.10d)$$

The present form of the $\Delta$ propagator in eq. (3.10a,b,c,d)   
is similar to the quantum mechanical form of the Breit-Wigner propagator,
where a imagenery part of energy of resonance is
generated by a retardation effect \cite{Gold}.

${\cal D}_{\gamma'\pi' N'-\pi N}^{\mu}$ (3.7a)
satisfies the same current conservation condition (3.3b) as
${\widetilde {\cal E}}_{\gamma'\pi' N'-\pi N}^{\mu}$ (2.15) and 
 ${\cal M}_{\gamma'\pi' N'-\pi N}^{\mu}$ (3.4a). 
Unlike the  ${\cal D}_{\gamma'\pi' N'-\pi N}^{\mu}$ amplitude
${\cal H}_{\gamma'\pi' N'-\pi N}^{\mu}$ 
contains projections on the 
intermediate spin $3/2$ states. Therefore
${\cal H}_{\gamma'\pi' N'-\pi N}^{\mu}$
satisfies a modified current conservation condition

$$
\Biggl[{k'}_{\mu}
<out;{\bf p'}_{N}{\bf p'}_{\pi}|J^{\mu}(0)|{\bf p}_{\pi}{\bf p}_N;in>
\Biggr]_{on\ mass\ shell\ \pi',\ N',\ \pi,\ N}
^{Projection\ on\ spin\ 3/2\ particle\  states}
={k'}_{\mu}{\cal H}_{\gamma'\pi' N'-\pi N}^{\mu}
+{\sl b}_{\gamma'\pi' N'-\pi N}
=0,\eqno(3.11a)$$

where

$${\sl b}_{\gamma'\pi' N'-\pi N}=
{{ (p'_N)_{a} (p_N)_d }\over{ 
(p'_N{\bf .}p_N) }}
{\overline u}({\bf p'_N})i\gamma_5u^a({\bf P'})\Bigl\{
{\overline u}^b({\bf P'})g_{bc}u^c({\bf P})\Bigr\}
{\overline u}^d({\bf P})i\gamma_5u({\bf p_N})$$
$$\Biggl[
{\overline u}({\bf p_N})
{{ (p'_N+k')_{\sigma}\gamma^{\sigma}+m_N}\over {2m_N}}
e_{N'}<out;{\bf p'}_{\pi}|J(0)|{\bf p}_{\pi}{\bf p}_N;in>$$
$$-e_{N}<out;{\bf p'}_{\pi}{\bf p'}_{N}|{\overline J}(0)|{\bf p}_{\pi};in>
{{ (p_N-k')_{\sigma}\gamma^{\sigma}+m_N}\over {2m_N}}u({\bf p'_N})$$
$$
+{\overline u}({\bf p_N})u({\bf p'_N})
e_{\pi'}<out;{\bf p'}_{N}|j_{\pi'}(0)|{\bf p}_{\pi}{\bf p}_N;in>
-e_{\pi}<out;{\bf p'}_{\pi}{\bf p'}_{N}|j_{\pi}(0)|{\bf p}_{N};in>
{\overline u}({\bf p_N})u({\bf p'_N})
\Biggr]\eqno(3.11b)$$
is obtained from ${\widetilde {\cal B}}_{\gamma'\pi' N'-\pi N}$ (2.14)
after the intermediate projection on the spin $3/2$ particle states.

Substitution of (3.10a,b,c,d) into (3.8) gives

$${\cal H}_{\gamma'\pi' N'-\pi N}^{\mu}=
{1\over{ (p'_N{\bf .}p_N) }}
{\overline u}({\bf p'_N})(p'_N)_{a}i\gamma_5u^a({\bf P'})$$
$$\Biggl\{
{\overline u}^b({\bf P'})g_{bc}\Bigl[
(P+P')^{\mu}{\cal V}_E-i\sigma^{\mu\nu}k'_{\nu}{\cal V}_H
\Bigr]u^c({\bf P})\Biggr\}
{\overline u}^d({\bf P})(p_N)_d i\gamma_5u({\bf p_N}),\eqno(3.12a)$$

where 

$${\cal V}_E=
{{ e_{N'}{\sc R}_{N'}}\over{(s-s')
\Bigl(
{p}_{\pi}^{o}+{p}_{N}^{o}-{P}^o_{\Delta}(s)\Bigr)}}
+{{e_{\pi'} {\sc R}_{\pi'}}
\over{(s-s')
\Bigl(
{p}_{\pi}^{o}+{p}_{N}^{o}-{P}^o_{\Delta}(s)\Bigr) }}$$
$$-{{e_{N}{\sc R}_{N}}
\over{(s-s')\Bigl(
{p'}_{\pi}^{o}+{p'}_{N}^{o}-{P'}^o_{\Delta}(s')\Bigr) }}
-{{e_{\pi} {\sc R}_{\pi} }\over{
(s-s')
\Bigl({p'}_{\pi}^{o}+{p'}_{N}^{o}-{P'}^o_{\Delta}(s')
\Bigr) }},
\eqno(3.13a)$$

$${\cal V}_H=
{{ \mu_{N'}{\sc R}_{N'}}\over{(s-s')
\Bigl( {p}_{\pi}^{o}+{p}_{N}^{o}-{P}^o_{\Delta}(s)\Bigr) }}
-{{\mu_{N}{\sc R}_{N}}
\over{(s-s')
\Bigl({p'}_{\pi}^{o}+{p'}_{N}^{o}-{P'}^o_{\Delta}(s')
\Bigr) }}.
\eqno(3.13b)$$

After a simple algebra (3.13b) can be rewritten as

$${\cal V}_H=
{1\over{
\Bigl({p'}_{\pi}^{o}+{p'}_{N}^{o}-{P'}^o_{\Delta}(s')\Bigr)
\Bigl({p}_{\pi}^{o}+{p}_{N}^{o}-{P}^o_{\Delta}(s)\Bigr) }}
\Biggl\{ R_+\biggl[
{{|{\bf k'}|}\over {s-s'}}
-{ {{P}^o_{\Delta}(s)-{P'}^o_{\Delta}(s')}\over{s-s'}}
\biggr]
$$
$$+{{R_-}\over {s-s'} }
\biggl[ 
{p}_{\pi}^{o}+{p_{N}}^{o}+{p'}_{\pi}^{o}+{p'_{N}}^{o}
-{P}^o_{\Delta}(s)-{P'}^o_{\Delta}(s')
\biggr]\Biggr\}
\eqno(3.14a)$$

where 
$$R_{\pm}
={1\over 2}\Bigl[\mu_{N'}{\sc R}_{N'}
\pm\mu_{N}{\sc R}_{N}\Bigr].\eqno(3.14b)$$

The first part in (3.14a) is regular at $|{\bf k'}|=0$, where
$s=s'$, because $\Bigl({P}^o_{\Delta}(s)-{P'}^o_{\Delta}(s')\Bigr)/(s-s')$
is  finite at $s=s'$. In the $\pi N$ c.m. frame by $|{\bf k'}|=0$ 
this part is in order of $~1/(\Gamma_{\Delta}/2)^2$, where
$\Gamma_{\Delta}$ is the $\Delta$ decay width.

The second part of ${\cal V}_H$ (3.14a) can describe only one 
$\Delta$ exchange because
${p}_{\pi}^{o}+{p_{N}}^{o}+{p'}_{\pi}^{o}+{p'_{N}}^{o}
-{P}^o_{\Delta}(s)-{P'}^o_{\Delta}(s')$ cancel with the one of the $\Delta$
propagators. This expression 
has a different behavior in two respects:

${\bf 1.}$ For the  $\pi N$ bremsstrahlung reactions without charge exchange
$\pi^{\pm} p\longrightarrow\gamma \pi^{\pm} p$ or 
$\pi^{o} p\longrightarrow\gamma \pi^{o} p$ 
we have $\mu_{N'}=\mu_N$. In this case     
$R_-/ (s-s')$ is finite at threshold $|{\bf k'}|=0$. 
In the $\Delta$ resonance region the second part of (3.14a) is in order of  
$~1/(\Gamma_{\Delta}/2)$.

${\bf 2.}$  For the charge exchange reactions $R_-/ (s-s')$ can be singular at 
threshold $|{\bf k'}|=0$. This case needs a special investigation.

In the complex energy region, where    
$p_{\pi}^{\mu}+p_{N}^{\mu}\Longrightarrow P^{\mu}_{\Delta}$
 and ${p'}_{\pi}^{\mu}+{p'}_{N}^{\mu}\Longrightarrow {P'}^{\mu}_{\Delta}$
both  parts of (3.14a) contain the same singularities in the propagators
${1/{\Bigl({p'}_{\pi}^{o}+{p'}_{N}^{o}-{P'}^o_{\Delta}(s')\Bigr)
\Bigl({p}_{\pi}^{o}+{p}_{N}^{o}-{P}^o_{\Delta}(s)\Bigr) }}$.  Therefore 
it is convenient to introduce the 
smooth function in the above limits  ${\cal V}_H$

$${\cal V}_H=
{1\over{
\Bigl({p'}_{\pi}^{o}+{p'}_{N}^{o}-{P'}^o_{\Delta}(s')\Bigr)
\Bigl({p}_{\pi}^{o}+{p}_{N}^{o}-{P}^o_{\Delta}(s)\Bigr) }}
{\rm g}_{\pi' N'-\Delta'}(s',k'){\sc V}_H
{\rm g}_{\Delta-\pi N}(s),
\eqno(3.15a)$$

where 
${\rm g}_{\Delta-\pi N}(s)$ and ${\rm g}_{\pi' N'-\Delta'}(s',k')$
are the radial parts of the $\Delta-\pi N$ form factors from
Fig. 2B.

The same transformations 
for ${\cal V}_E$ gives

$${\cal V}_E=
{1\over{
\Bigl({p'}_{\pi}^{o}+{p'}_{N}^{o}-{P'}^o_{\Delta}(s')\Bigr)
\Bigl({p}_{\pi}^{o}+{p}_{N}^{o}-{P}^o_{\Delta}(s)\Bigr) }}
{\rm g}_{\pi' N'-\Delta'}(s',k'){\sc V}_E
{\rm g}_{\Delta-\pi N}(s).
\eqno(3.15b)$$

Relations (3.15a,b) enables us to represent external particle 
radiation amplitude (3.12a) in the form of the double $\Delta$
particle exchange amplitude (Fig. 2B)

$$
{\cal H}_{\gamma'\pi' N'-\pi N}^{\mu}=
{-1\over{(p'_N{\bf .}p_N) }}
{{<{\bf p'}_N,{\bf p'}_{\pi}|{\sl g}_{\pi' N'-\Delta'}|{\bf P'}_{\Delta}>}
\over{{p'}_{\pi}^{o}+{p'}_{N}^{o}-{P'}^o_{\Delta}(s')}}$$
$$\Biggl\{
{\overline u}^b({\bf P'})g_{bc}\Bigl[
(P+P')^{\mu}{\sc V}_E-i\sigma^{\mu\nu}k'_{\nu}{\sc V}_H
\Bigr]u^c({\bf P})\Biggr\}
{{<{\bf P}_{\Delta}|{\sl g}_{\Delta-\pi N}|{\bf p}_N,{\bf p}_{\pi}>}
\over{{p}_{\pi}^{o}+{p}_{N}^{o}-{P}^o_{\Delta}(s) }},
\eqno(3.12b)$$

where the following $\Delta-\pi N$ and $\pi N-\Delta$ vertex functions  
are introduced{\footnotemark}

\footnotetext{ 
For the on mass shell $\pi N$ states with
$s=s'$ and $k'=0$ the operator

$$ {\cal Q}({\bf p'}_N,{\bf p}_N,{\bf P'}_{\Delta},{\bf P}_{\Delta})=
{1\over{(p'_N{\bf .}p_N) }}
{\overline u}({\bf p'}_N)i\gamma_5({{p'}_N})_a u^a({\bf P'}_{\Delta})
{\overline u}^d({\bf P}_{\Delta})({{p}_N})_d i\gamma_5 u({\bf p}_N)$$
transforms into the projection operator ${\cal P}_1^{3/2}({\bf p'}_N,{\bf p}_N)$
which projects on the $\pi N$ state with 
the orbital momentum $L=1$ and the total momentum $J=3/2$ \cite{MF} 

$${\cal P}_1^{3/2}({\bf p'}_N,{\bf p}_N)=
{{6m_N}\over{4\pi{\bf p}{\bf p'}(m_N+\sqrt{m_N^2+{\bf p}^2})}})
{\overline u}({\bf p'}_N)i\gamma_5{{p'}_N}_a u^a({\bf P'}_{\Delta})
{\overline u}^d({\bf P}_{\Delta}){{p}_N}_d i\gamma_5 u({\bf p}_N),$$

i.e.

$$\Biggl[{\cal Q}({\bf p'}_N,{\bf p}_N,{\bf P'}_{\Delta},{\bf P}_{\Delta})
\Biggr]^{|{\bf k'}|=0}
= {{ 4\pi{\bf p}{\bf p'}(m_N+\sqrt{m_N^2+{\bf p}^2} ) }
\over{6m_N (p'_N{\bf .}p_N) }}
{\cal P}_1^{3/2}({\bf p'}_N,{\bf p}_N)$$
}

$$<{\bf p'}_N,{\bf p'}_{\pi}|{\sl g}_{\pi' N'-\Delta}|{\bf P'}_{\Delta}>=
{\rm g}_{\pi' N'-\Delta'}(s',k')
{\overline u}({\bf p'}_N)i\gamma_5{{p'}_N}_a u^a({\bf P'}_{\Delta}),
\eqno(3.16a)$$

$$<{\bf P}_{\Delta}|{\sl g}_{\Delta-\pi N}|{\bf p}_N,{\bf p}_{\pi}>=
{\rm g}_{\Delta-\pi N}(s)
{\overline u}^d({\bf P}_{\Delta}){{p}_N}_d i\gamma_5 u({\bf p}_N).
\eqno(3.16b)$$

The total $\pi N$ bremsstrahlung amplitude
$<out;{\bf p'}_{N}{\bf p'}_{\pi}|J^{\mu}(0)|{\bf p}_{\pi}{\bf p}_N;in>$
is a sum of the photon radiation diagrams from 
the external particles  (2.8b) (Fig. 1) and the infinite  set of
other amplitudes with  
the photon radiation diagrams from the internal particles 
${\cal I}_{\gamma'\pi' N'-\pi N}^{\mu}$. A symbolical
picture of this internal amplitude is given  on Fig. 2A.

$$<out;{\bf p'}_{N}{\bf p'}_{\pi}|J^{\mu}(0)|{\bf p}_{\pi}{\bf p}_N;in>
={\cal E}_{\gamma'\pi' N'-\pi N}^{\mu}+
{\cal I}_{\gamma'\pi' N'-\pi N}^{\mu}.
\eqno(3.17a)$$
The current conservation allows to use
the well known Low prescription \cite{Low} for an estimation
of ${\cal I}_{\gamma'\pi' N'-\pi N}^{\mu}$ on the basis of the external
particle radiation amplitude.
In particular, from the current conservation (3.11a)
follows the existence of  an internal particle radiation amplitude
${\sc I}_{\gamma'\pi' N'-\pi N}^{\mu}$
which satisfies the condition

$${k'}_{\mu}{\widetilde {\cal I}}_{\gamma'\pi' N'-\pi N}^{\mu}= 
{\sl b}_{\gamma'\pi' N'-\pi N},
\eqno(3.17b)$$
where  the intermediate antinucleon degrees of freedom are omitted.

From the infinite set of the internal particle radiation diagrams 
we take the double $\Delta$ exchange term
${\sc H}_{\gamma'\pi' N'-\pi N}^{\mu}$  {\footnotemark}  (Fig. 2B)
which has the same analytical properties as
 ${\cal H}_{\gamma'\pi' N'-\pi N}^{\mu}$ (3.12b).
This term contains   
the full $\Delta -\gamma'\Delta'$ 
vertex function $<{\bf P'}_{\Delta}|J_{\mu}(0)|{\bf P}_{\Delta}>$.

$${\sc H}_{\gamma'\pi' N'-\pi N}^{\mu}=
{-1\over{(p'_N{\bf .}p_N) }}
{{<{\bf p'}_N,{\bf p'}_{\pi}|{\sl g}_{\pi' N'-\Delta'}|{\bf P'}_{\Delta}>}
\over{{p'}_{\pi}^{o}+{p'}_{N}^{o}-{P'}^o_{\Delta}(s')}}
<{\bf P'}_{\Delta}|J^{\mu}(0)|{\bf P}_{\Delta}>
{{<{\bf P}_{\Delta}|{\sl g}_{\Delta-\pi N}|{\bf p}_N,{\bf p}_{\pi}>}
\over{{p}_{\pi}^{o}+{p}_{N}^{o}-{P}^o_{\Delta}(s) }}.
\eqno(3.18a)$$

This expression can be exactly 
extracted from the $\pi N$ bremsstrahlung amplitude 
as it was done
for the amplitude  of the  $\gamma p\to\gamma'\pi'N'$ reaction
in our previous papers \cite{Ann,MF}. Corresponding recipe
is given in  appendix B. A connection between the external particle 
radiation diagrams on Fig. 1 and double $\Delta$ exchange term on Fig. 2
in tree approximation are considered in \cite{Ding} using the Brodsky-Brown 
decomposition identities \cite{Brodsky1,Brodsky2}.

\footnotetext{ 
An other double $\Delta$ exchange term
contains the $\Delta-\pi'\Delta'$ vertex function.
This term has  a different analytical 
behavior as (3.12b) and (3.18a) with $\Delta-\gamma'\Delta'$ vertex function.
In addition, the term with $\Delta-\pi'\Delta'$ vertex is small \cite{Mink}. }

 The general form of the $\Delta -\gamma'\Delta'$ vertex is

$$<{\bf P'}_{\Delta}|J^{\mu}(0)|{\bf P}_{\Delta}>
=(P_{\Delta}+P'_{\Delta})^{\mu}\Biggl(
{\overline u}^{\sigma}({\bf P'}_{\Delta})
\Bigl[g_{\rho\sigma}G_1({k'}_{\Delta}^2)
+{k'_{\Delta}}_{\sigma}{k'_{\Delta} }_{\rho}G_3({k'}_{\Delta}^2)
\Bigr]\Biggr)u^{\rho}({\bf P}_{\Delta})$$
$$+{\overline u}^{\sigma}({\bf P'}_{\Delta})\Biggl(
-i\sigma^{\mu\nu}{k'_{\Delta}}_{\nu}\Bigl[g_{\rho\sigma}G_2({k'}_{\Delta}^2)
+{k'_{\Delta}}_{\sigma}{k'_{\Delta} }_{\rho}G_4({k'}_{\Delta}^2)
\Bigr]\Biggr)u^{\rho}({\bf P}_{\Delta}),\eqno(3.19a)$$

where ${k'_{\Delta}}_{\mu}=(P_{\Delta}-P'_{\Delta})_{\mu}$,
$g_{\mu\nu}$ is the metric tensor and  
the form factors $G_i({k'}_{\Delta}^2)$ are simply connected with 
the charge monopole
$G_{C0}({k'}_{\Delta}^2)$, the magnetic dipole $G_{M1}({k'}_{\Delta}^2)$, 
the electric quadrapole
$G_{E2}({k'}_{\Delta}^2)$ and the magnetic octupole
$G_{M3}({k'}_{\Delta}^2)$ form factors of the $\Delta$ resonance.
{\footnotemark}
In the low energy region we can neglect the terms 
$\sim {k'}_{\Delta}^2/4M_{\Delta}^2$,
and keep only terms $\sim 1/M_{\Delta}$. Then the previous formula
can be rewritten in a similar form as the $\gamma N N$ vertex
function:

\footnotetext{ 
Other choices of $F_i$ form factors are considered in ref. \cite{PasVan3} }

$$<{\bf P'}_{\Delta}|J_{\mu}(0)|{\bf P}_{\Delta}>=
{\overline u}^{\sigma}({\bf P'}_{\Delta})g_{\rho\sigma}\biggl[ 
{{(P_{\Delta}+P'_{\Delta})_{\mu}}\over{2M_{\Delta}} }G_{C0}({k'}_{\Delta}^2)- 
{{i\sigma_{\mu\nu}{k'}_{\Delta}^{\nu}}\over{2M_{\Delta} }}
G_{M1}({k'}_{\Delta}^2)
\biggr]u^{\rho}({\bf P}_{\Delta})\eqno(3.19b)$$
The form of the expressions (3.19a,b) insures the validity of the
the one-body current conservation  in the complex 
$\Delta$-resonance energy region

$${k'}_{\Delta}^{\mu}<{\bf P'}_{\Delta}|J_{\mu}(0)|{\bf P}_{\Delta}>=0,\ \ \
 {{k'}_{\Delta}}_{\mu}{\sc H}_{\gamma'\pi' N'-\pi N}^{\mu}=0.\eqno(3.18b)$$

Thus after projections on the spin $3/2$ states 
in the full $\pi N$ bremsstrahlung amplitude 
$<out;{\bf p'}_{N}{\bf p'}_{\pi}|J^{\mu}(0)|{\bf p}_{\pi}{\bf p}_N;in>$ (3.17a)
we have separated two leading amplitudes (3.12b) and (3.18a)
with the double $\Delta$-exchange poles and $\Delta-\Delta'\gamma'$ vertex.
Unfortunately ${\cal H}_{\gamma'\pi' N'-\pi N}^{\mu}$ (3.12b)
contains also 
a contributions from the one $\Delta$ exchange (see consideration after eq. 
(3.14b)). This one can see from the structure of
 ${\cal H}_{\gamma'\pi' N'-\pi N}^{\mu}$ which
does not satisfy the one body current conservation condition 
${{k'}_{\Delta}}_{\mu}{\cal H}_{\gamma'\pi' N'-\pi N}^{\mu}\ne 0$.  
Therefore in order to compare ${\cal H}_{\gamma'\pi' N'-\pi N}^{\mu}$
with ${\sc H}_{\gamma'\pi' N'-\pi N}^{\mu}$
we must
modify ${\cal H}_{\gamma'\pi' N'-\pi N}^{\mu}$ as
$${\cal H}_{\gamma'\pi' N'-\pi N}^{\mu}\longrightarrow
{\widetilde {\cal H}}_{\gamma'\pi' N'-\pi N}^{\mu},\ \ \  
{\widetilde {\cal H}}_{\gamma'\pi' N'-\pi N}^{\mu}=
{\cal H}_{\gamma'\pi' N'-\pi N}^{\mu}-
{\sl h}_{\gamma'\pi' N'-\pi N}^{\mu},\eqno(3.20)$$

where

$${\widetilde {\cal H}}_{\gamma'\pi' N'-\pi N}^{\mu}=
{-1\over{(p'_N{\bf .}p_N) }}
{{<{\bf p'}_N,{\bf p'}_{\pi}|{\sl g}_{\pi' N'-\Delta'}|{\bf P'}_{\Delta}>}
\over{{p'}_{\pi}^{o}+{p'}_{N}^{o}-{P'}^o_{\Delta}(s')}}$$
$$\Biggl\{
{\overline u}^b({\bf P'})g_{bc}\Bigl[
(P_{\Delta}+P'_{\Delta})^{\mu}{\sc V}_E-
i\sigma^{\mu\nu}{k'_{\Delta}}_{\nu}{\sc V}_H\Bigr]u^c({\bf P})\Biggr\}
{{<{\bf P}_{\Delta}|{\sl g}_{\Delta-\pi N}|{\bf p}_N,{\bf p}_{\pi}>}
\over{{p}_{\pi}^{o}+{p}_{N}^{o}-{P}^o_{\Delta}(s) }},
\eqno(3.21a)$$

$${\sl h}_{\gamma'\pi' N'-\pi N}^{\mu}=
{-1\over{(p'_N{\bf .}p_N) }}
{{<{\bf p'}_N,{\bf p'}_{\pi}|{\sl g}_{\pi' N'-\Delta'}|{\bf P'}_{\Delta}>}
\over{{p'}_{\pi}^{o}+{p'}_{N}^{o}-{P'}^o_{\Delta}(s')}}$$
$$\Biggl\{
{\overline u}^b({\bf P'})g_{bc}\Bigl[
(P_{\Delta}+P'_{\Delta}-P-P')^{\mu}{\sc V}_E-
i\sigma^{\mu\nu}
(k'_{\Delta}-k')_{\nu}{\sc V}_H\Bigr]u^c({\bf P})\Biggr\}
{{<{\bf P}_{\Delta}|{\sl g}_{\Delta-\pi N}|{\bf p}_N,{\bf p}_{\pi}>}
\over{{p}_{\pi}^{o}+{p}_{N}^{o}-{P}^o_{\Delta}(s) }},
\eqno(3.21b)$$

Expression  (3.21b) contains  factors $(P_{\Delta}+P'_{\Delta}-P-P')$
and $k'_{\Delta}-k'$ which cancel with one of the $\Delta$ propagators
$1/({P'}^{o}-{P'}^o_{\Delta})$ or $1/(P^{o}-{P}^o_{\Delta})$. Therefore
${\sl h}_{\gamma'\pi' N'-\pi N}^{\mu}$ contributes to the one $\Delta$ 
exchange part of external particle radiation amplitude.
This contribution is in order of $2/\Gamma_{\Delta}$ unlike to 
the double $\Delta$ exchange terms    
$\sim 4/\Gamma_{\Delta}^2$. Therefore if we keep only the double 
$\Delta$ exchange terms, then
${\widetilde {\cal H}}_{\gamma'\pi' N'-\pi N}^{\mu}$
satisfies the one-body   current conservation condition

$${{k'}_{\Delta}}_{\mu}
{\widetilde {\cal H}}_{\gamma'\pi' N'-\pi N}^{\mu}=0\eqno(3.22)$$
  
and instead of (3.17b) we get the following
real photon current conservation condition

$${{k'}}_{\mu}\Bigl(
{\widetilde {\cal H}}_{\gamma'\pi' N'-\pi N}^{\mu}+
{\sc H}_{\gamma'\pi' N'-\pi N}^{\mu}\Bigr)=0.\eqno(3.23)$$

It is evident, that relation (3.23) implies a appropriate modification of 
(3.17b) 

$$k'_{\mu}{\widetilde {\cal H}}_{\gamma'\pi' N'-\pi N}^{\mu}
={\widetilde {\sl b}}_{\gamma'\pi' N'-\pi N},\eqno(3.17c)$$

where instead of (3.11b) we get

$$
{\widetilde{\sc b}}_{\gamma'\pi' N'-\pi N}^{\mu}=
{-1\over{(p'_N{\bf .}p_N) }}
{{<{\bf p'}_N,{\bf p'}_{\pi}|{\sl g}_{\pi' N'-\Delta'}|{\bf P'}_{\Delta}>}
\over{{p'}_{\pi}^{o}+{p'}_{N}^{o}-{P'}^o_{\Delta}(s')}}$$
$$\Biggl\{
{\overline u}^b({\bf P'})g_{bc}\Bigl[
(P_{\Delta}+P'_{\Delta})^{\mu}{\sc V}_E-i\sigma^{\mu\nu}
{k'_{\Delta}}_{\nu}{\sc V}_H
\Bigr]u^c({\bf P})\Biggr\}
{{<{\bf P}_{\Delta}|{\sl g}_{\Delta-\pi N}|{\bf p}_N,{\bf p}_{\pi}>}
\over{{p}_{\pi}^{o}+{p}_{N}^{o}-{P}^o_{\Delta}(s) }},
\eqno(3.11c)$$

 The importance of eq. (3.23) follows from the cancellation 
of the internal $ {\sc H}_{\gamma'\pi' N'-\pi N}^{\mu}$
and external ${\widetilde {\cal H}}_{\gamma'\pi' N'-\pi N}^{\mu}$ amplitudes
according to  current conservation. 
Thus according to (3.23) a comparison of the $\Delta-\gamma'\Delta'$ 
vertex functions 
in (3.21a) and (3.18a)  gives   

$${\overline u}^{\sigma}({\bf P'}_{\Delta})g_{\rho\sigma}\biggl[
{{k'_{\mu}(P_{\Delta}+P'_{\Delta})^{\mu}}\over{2M_{\Delta}} }
(2M_{\Delta}{\sc V}_E)-
{{ik'_{\mu}\sigma_{\mu\nu}{k'}_{\Delta}^{\nu}}\over{2M_{\Delta} }}
(2M_{\Delta}{\sc V}_H)\biggr]u^{\rho}({\bf P}_{\Delta})$$
$$=-{\overline u}^{\sigma}({\bf P'}_{\Delta})g_{\rho\sigma}\biggl[ 
{{k'_{\mu}(P_{\Delta}+P'_{\Delta})^{\mu}}\over{2M_{\Delta}} }
G_{C0}({k'}_{\Delta}^2)- 
{{ik'_{\mu}\sigma_{\mu\nu}{k'}_{\Delta}^{\nu}}\over{2M_{\Delta} }}
G_{M1}({k'}_{\Delta}^2)\biggr]
u^{\rho}({\bf P}_{\Delta}),\eqno(3.24)$$

where we have used a low energy photon approximation 
which enables us to ignore 
the electric quadrupole and the magnetic octupole parts of the 
$\Delta-\gamma\Delta$ vertex functions. 
Here it's important to note, that
$2M_{\Delta}$ is a natural unit for 
${\sc V}_E$ and ${\sc V}_H$, where $2M_{\Delta}/(s-s')$ 
can be replaced in $\Delta$ resonance region by the linear propagator 
$\sim 1/(s^{1/2}-{s'}^{1/2})$ which is used in the considered 
time-ordered formulation.

Next from (3.24) we get 
$$G_{C0}({k'}_{\Delta}^2)= -2M_{\Delta}{\sc V}_E;\ \ \ 
G_{M1}({k'}_{\Delta}^2)=-2M_{\Delta}{\sc V}_H,\eqno(3.25)$$
where ${\sc V}_E$ and ${\sc V}_H$, as well as ${\sc r}_{N'}$,
${\sc r}_{\pi'}$, ${\sc r}_{N}$ and ${\sc r}_{\pi}$ in (3.10a,b,c,d)
are given in the limits
$$p_{\pi}^{\mu}+p_{N}^{\mu}\Longrightarrow P^{\mu}_{\Delta},\ \ \ and\ \ \ 
{p'}_{\pi}^{\mu}+{p'}_{N}^{\mu}\Longrightarrow {P'}^{\mu}_{\Delta}.
\eqno(3.26a)$$

Here a special on mass shell 
momenta of $\pi$ mesons 
($q_{\pi}=(\sqrt{ {\bf q}_{\pi}^2+m_{\pi}^2},{\bf q}_{\pi})$,
${q'}_{\pi}=(\sqrt{ {\bf q'}_{\pi}^2+m_{\pi}^2},{\bf q'}_{\pi})$ 
and of nucleons
($q_{N}=(\sqrt{ {\bf q}_{N}^2+m_{N}^2},{\bf q}_{N})$,
${q'}_{N}=(\sqrt{ {\bf q'}_{N}^2+m_{N}^2},{\bf q'}_{N})$ 
are introduced.
These four-momenta have the equal radial part and the same direction as the 
corresponding on-mass shell four momenta ${\bf p}$  

$${\bf q}_N=q { {{\bf p}_N}\over
{p_N}},\ \ \ {\bf q}_{\pi}=q {{{\bf p}_{\pi}}\over
{p_{\pi}}},
\ \ \ 
{\bf q'}_N=q { {{\bf p'}_N}\over {p'_N}},\ \ \ {\bf q'}_{\pi}=q 
{{ {\bf p'}_{\pi}}\over {p'_{\pi}}},\eqno(3.26b)$$

where

$$q^2= { {\Bigl(m^2_{\Delta}-(m_N+m_{\pi})^2\Bigr)
\Bigl(m^2_{\Delta}-(m_N-m_{\pi})^2\Bigr) }
\over{4m_{\Delta}^2 }}.\eqno(3.26c)$$

In ${\sc V}_E$ and ${\sc V}_H$  
 integrations are implied over the 
relative $\pi$-$N$ angles  
due to projection on the states with the spin $3/2$
and $L=1$ orbital momenta. Finally ${\sc V}_E$ and ${\sc V}_H$
are dependent only on ${k'}_{\Delta}^2$.
It must be emphasized, that the Ward-Takahashi identity 
(2.4) remains to be valid for the bremsstrahlung amplitude
in the limits $\lim_{p_{\pi}^{o}+p_{N}^{o}\Longrightarrow P^{o}_{\Delta}}
\lim_{{p'}_{\pi}^{o}+{p'}_{N}^{o}\Longrightarrow {P'}^{o}_{\Delta}}
  <out;{\bf p'}_{N}{\bf p'}_{\pi}|J^{\mu}(0)|{\bf p}_{\pi}{\bf p}_N;in>$,
because ${ q}_N$, ${ q}_{\pi}$, ${ q'}_N$ and ${ q'}_{\pi}$ are on mass shell.

According to (3.25) the electric  and magnetic parts 
 of the internal and external particle radiation amplitudes 
with double $\Delta$ exchange and $\Delta-\Delta\gamma$ vertex
cancel each other. Therefore the full 
$\pi N$ bremsstrahlung  amplitude become independent on these amplitudes.
Thus the role of the double $\Delta$ exchange amplitude
reduces to decrease the external particle radiation parts
of the $\pi N$ bremsstrahlung amplitude.
Consequently we have screening of the internal $\Delta$ radiation part by
the external particle radiation diagrams.

Equation (3.25) indicates that for ${\sc V}_E$ (3.15a)  a 
normalization condition is fulfilled for ${\sc V}_E$ and $G_{C0}$

$$G_{C0}(0)=-\biggl[{2M_{\Delta}\sc V}_E\biggr]^{k'=0}=e_{\Delta}\eqno(3.27)$$
where $e_{\Delta}$ denotes the charge of $\Delta$. 

The normalization condition (3.27)
can be used also for ${\sc V}_H$ in the special 
cases for neutral pion  reactions $\pi^o p\to\gamma' {\pi^o}' p'$ and    
$\gamma p\to\gamma' \pi^o p'$, where ${\sc V}_E$ and ${\sc V}_H$
contains only nucleon parts and 
they are determined  by the same expression in the square brackets of eq.
(3.8). Then according to (3.27) it is easy get

$$\mu_{\Delta^+}=G_{M1}(0)=-\biggl[{2M_{\Delta}\sc V}_H\biggr]^{k'=0}=
\mu_p{ {M_{\Delta} }\over {m_N} }\eqno(3.28)$$

Magnetic momenta of $\Delta^{++}$ can be obtained from (3.28) using 
the isotopic 
symmetry between the $\pi^o p\to \pi^o p$ and $\pi^+ p\to \pi^+ p$
 amplitudes. Then we get  
$\mu_{\Delta^{++}}={3\over 2}\mu_{\Delta^+}$. Unfortunately one can not use
the $\pi^o n\to\gamma' {\pi^o}' n'$ reaction for determination of 
the magnetic moments of $\Delta^{o}$ and $\Delta^{-}$, because
equal-time commutators (2.3a,b) are zero in that case.

The present procedure of construction of the bremsstrahlung amplitude does not 
contradict  to the 
quantum-electro-dynamical (QED) renormalisation recipe of the charge and 
the mass. 
Thus if we assume that the perturbation series  is convergent, then we
can operate with the physical charge and the physical mass  
which enter
in the above relation for the equal time commutators and  in the corresponding 
amplitudes. Moreover, the renormalisation of the already renormalized
expressions generates additional conditions \cite{M83} which preserve
the form of the equal-time commutators and other principal  conditions.

\vspace{0.25cm}

\begin{center}
                  {\bf 4. Conclusion}
\end{center}

\vspace{0.25cm}

As basis for the study of the $\pi N$ bremsstrahlung we have used a two-body
form of the Ward-Takahashi identities (2.4) which generates following
model-independent relations:

${\bf (i)}$ An amplitude of an arbitrary 
$a+b\longrightarrow \gamma'+f_1+...+f_n$ $(n=1,2,...)$ 
reaction fulfill  generalized current
conservations

$$k'_{\mu}<out;f_1,...,f_n|J^{\mu}(0)|a,b;in>=
\Biggl[{\cal B}_{\gamma'f_1...f_n-a b}+
{k'}_{\mu}{\cal E}_{\gamma'f_1..f_n-a b}^{\mu}
\Biggr]_{on\ mass\ shell\ f_1,\ .\ .\ .\  f_n;\ a,\ b}=0,\eqno(I)$$

where 
${\cal E}_{\gamma'f_1..f_n-a b}^{\mu}$ corresponds to the
complete set of Feynman (or three-dimensional time-ordered) diagrams with 
the photo-emission from each external particles and

$${\cal B}_{\gamma'f_1..f_n-a b}=
\sum_{m=1(I_1\ne m...I_{n-1}\ne m)}^n e_m<out; f_{I_1}...f_{I_{n-1}}|
J_m(0)|a,b;in>$$
$$-e_a<out; f_1...f_n|J_a(0)|b;in>-
e_b<out; f_1...f_n|J_b(0)|a;in>\eqno(II)$$
stands for amplitudes of  the $a+b\longrightarrow f_1+...+f_n$ reaction
without $\gamma'$ emission. 

A special case of relation (I)
is given by eq. (2.7a,b) with the external particle radiation 
diagrams on Fig. 1.

Equation (I) and (II) are also valid for an arbitrary number of 
external photons. For instance, these equations can be used as the current 
conservation conditions  for the pion photo-production reaction 
$\gamma A\to \pi'A'$, for Compton scattering $\gamma A\to \gamma'A'$ etc.  
Moreover the
current conservation condition (I) can be extended in the off mass shell region
as it is done in eq. (2.7b).

${\bf (ii)}$ The current 
conservation (I) requires the  existence of the internal particle 
radiation amplitude ${\cal I}_{\gamma'f_1..f_n-a b}^{\mu}$ which ensures the 
validity of this condition  

$${k'}_{\mu}{\cal I}_{\gamma'f_1..f_n-a b}^{\mu}
={\cal B}_{\gamma'f_1...f_n-a b},\ or\ 
{k'}_{\mu}{\cal E}_{\gamma'f_1..f_n-a b}^{\mu}+
{k'}_{\mu}{\cal I}_{\gamma'f_1..f_n-a b}^{\mu}=0.
\eqno(III)$$

This means that ${\cal E}_{\gamma'f_1..f_n-a b}^{\mu}$ and 
${\cal I}_{\gamma'f_1..f_n-a b}^{\mu}$ have a different sign and they must be
subtracted from each other. Thus we have a screening  of the internal particle 
radiation amplitudes by the external one-particle radiation terms.

${\bf (iii)}$ For the soft emitted photons $k'\to 0$ our approach immediately
 reproduces the low energy theorems for the bremsstrahlung reactions.

${\bf (iv)}$
The external particle radiation part of the bremsstrahlung 
amplitude ${\cal E}^{\mu}$ contains the electromagnetic form factors
of the external particles only in the tree approximation. 
This follows from the equal-time commutators (2.3a,b) which are a 
result of the charge conservation. 
Thus we must modify the equal-time commutators (2.3a,b) between
the Heisenberg operators of the external particles
in order to apply the full electromagnetic form factors
of pions and nucleons in the current conservation condition (I) or (III).

The above screening mechanism has been applied to the
$\pi N$ bremsstrahlung reaction  with the leading double $\Delta$ 
exchange term (Fig. 2B). 
We have shown, that in the low energy region, where the electric quadruple 
and the magnetic octupole momenta of $\Delta$ can be neglected,
the above double $\Delta$ exchange term  is completely canceled 
with the corresponding part of the external particle radiation amplitude. 
From this cancellation follows the normalization 
condition  (3.25) for the Coulomb monopole part of the 
$\Delta-\gamma'\Delta'$ vertex which allows us to extract
the $\Delta^+$ and $\Delta^{++}$ dipole magnetic momenta   
$\mu_{\Delta^+}=G_{M1}(0)={ {M_{\Delta} }\over {m_N} }\mu_p$ and 
$\mu_{\Delta^{++}}={3\over 2}\mu_{\Delta^+}=5.46e/2m_p$ or 
$\mu_{\Delta^{++}}/\mu_p\sim 1.95$. Our result
for $\mu_{\Delta^{++}}$,
based on the model independent current conservation condition,   
is in agreement with the prediction of the naive $SU(6)$ quark 
model for $\mu_{\Delta^{++}}=2\mu_p=5.58e/2m_p$ \cite{Beg,Georgi},
with the nonrelativistic potential model \cite{Boss}
$\mu_{\Delta^{++}}=4.6\pm0.3$.
and with  extraction of $\mu_{\Delta^{++}}$ from the 
$\pi ^{+}p\to\gamma\pi^{+}p$ experimental cross section
in the framework of the low energy photon approach
 $\mu_{\Delta^{++}}=3.6\pm2.0$ \cite{Musa}, 
$\mu_{\Delta^{++}}=5.6\pm2.1$ \cite{Pascual} and 
$\mu_{\Delta^{++}}=4.7-6.9$\cite{Nefkens}. 
Our result is larger as the predictions in the modified 
$SU(6)$ models \cite{Brown,Pais} and 
in the soft-photon approximation 
$\mu_{\Delta^{++}}=3.7\sim 4.9 e/2m_p$ \cite{Lin}.
On the other hand
our result is smaller as the values obtained in the framework of the 
effective meson-nucleon Lagrangian  
$\mu_{\Delta^{++}}=6.1\pm 0.5 e/2m_p$ \cite{Castro},
 in the effective quark model $\mu_{\Delta^{++}}=6.17e/2m_p$ \cite{Franklin}
and in the modified bag model $\mu_{\Delta^{++}}=6.54$ \cite{Kriv}.

The summary of the numerical estimations of the magnetic moments of
$\Delta^+$ and $\Delta^{++}$ resonances is given in table 1. 
In a number of approaches the magnetic moment of $\Delta$
is treated as an adjustable parameter in the radiative $\pi N$
scattering which is determined using the most sensitive 
configurations to the $\Delta-\gamma\Delta$ vertex 
in the slow photon regime. Corresponding results obtained
 from the experimental cross sections of 
the $\pi^+p\to\gamma\pi^+p$ reaction are indicated 
in the table 1 with the  index $f$.
It must be emphasized, that only our approach and naive $SU(6)$ quark model
gives an analytical form for $\mu_{\Delta^{+}}$ and $\mu_{\Delta^{++}}$.
But our result for $\mu_{\Delta^{+}}$ is $M_{\Delta}/m_p\sim 1.31$-times
larger as  
$\mu_{\Delta^{+}}=\mu_{p}=2.79e/2m_p$ in refs. \cite{Beg,Franklin}.

\vspace{0.7cm}

\centerline{\bf Table  1}

\vspace{0.5cm}

\begin{center}
{\em Magnetic moments of $\Delta^+$ and $\Delta^{++}$
 in units of the nuclear magneton $\mu_N={e/{2m_N}}$. 
 The ref. in front of the index f indicates  the theoretical model  
  which is used to fit of the experimental data and to extract the 
magnetic moment   $\mu_{\Delta}$. }   
\end{center}
\hspace{-0.75cm}
\begin{tabular}{|c|c|c|c|c|c|c|c|} \hline\hline
{\sc Models}&      {\em This}               &  {\em $SU(6)$ }
            &      {\em Potential and}      &  {\em Modified}
            &      {\em Soft photon}         &  {\em Eff. $\pi N$ }
            &      {\em Eff. }               
\\
            &      {\em  work }             &
            &      {\em K-matrix appr.}     & {\em Bag}      
            &      {\em theorem  }          &  {\em Lagran.}      
            &      {\em quark   }         

\\    \hline
$\mu_{\Delta^+}$                 & 3.64     & 2.79 \cite{Beg,Georgi}
                                 &          &
                                 &          &
                                 & 2.79\cite{Franklin}

\\    \hline

                       &                          & 5.58 \cite{Beg,Georgi}
                       &6.9-9.7\cite{Heller}f     & 
                       &3.6$\pm$2.0\cite{Musa}f   &
                       &         
                                 
\\

${ \mu_{\Delta^{++}}}$ & 5.46                      &  4.25\cite{Pais} 
                       & 4.6$\pm$0.3\cite{Boss}f   &  6.54\cite{Kriv}
                       & 5.6$\pm$2.1\cite{Pascual}f&  6.1$\pm$0.5\cite{Castro}f
                       & 6.17\cite{Franklin}

\\
                       &                             & 4.41-4.89\cite{Brown}
                       &  5.6-7.5\cite{Wittman}f     &
                       &  4.7-6.9\cite{Nefkens}f     &
                       &          

\\                     &                             & 
                       &                             &
                       &  3.7-4.9\cite{Liou}f     &
                       &

\\ \hline \hline

\end{tabular}

\vspace{0.5cm}


This  screening 
mechanism can be observed in the cross sections of the $\pi N$ bremsstrahlung 
reaction or in the $\gamma p\to \gamma \pi^o p$ reaction
by comparison of 
the cross sections in and outside the $\Delta$ resonance region.
Due to the importance of the double $\Delta$ exchange diagram (Fig. 2B)
one must have a different  $1/k'$ behavior of the bremsstrahlung amplitude
in and outside the $\Delta$ resonance region.

Our  approach is based on current conservation  
in the  usual local quantum theory\cite{BD,IZ}. 
This approach is not dependent on the form of the Lagrangians or on the 
choice of the model of the $\pi N$ amplitudes i.e.
it is model independent. But 
this formulation does not include
the quark degrees of freedom which  violate the basic equal-time
commutators (2.3a,b) for the Ward-Takahashi identity. Therefore deviation
of our result from the experimental 
magnetic moments of $\Delta$-s can indicate the 
role  of  quark degrees of freedom in the electromagnetic interaction of
$\Delta$.

In classical electrodynamic it is known, that the current generated 
by acceleration of the charge $e$ is
$J_{\mu}=ie\biggl( {{p'_{\mu}}/     {(k'p')}}-{{p_{\mu}}/     {(k'p)}}
\biggr)$. Consequently by acceleration of two charged objects $e_1$
and $e_2$ we get a current 
$J^{\mu}=ie\biggl({p'_1}^{\mu}/(k'p'_1)V(1'.2')+{p'_2}^{\mu}/(k'p'_2)V(2'.1')
-p_1^{\mu}/(k'p_2)V(1,2)-p_2^{\mu}/(k'p_2)V(2,1)\biggr)$, where
$V(1,2)$ indicates a interaction between particle 1 and 2. 
Thus in classical electrodynamic the form of the currents is the same 
as in quantum field theory.
Therefore the external particle radiation amplitude (2.8b) and (2.15) can be 
interpreted as the sum of external pion and nucleon one-body currents
with the quantum mechanical $\pi N$ corrections.
One can hope, that the screening mechanism works also in classical 
electrodynamic.

\newpage

\begin{center}
                  {{\bf Appendix A:} Projections on the intermediate spin 
$3/2$ states}
\end{center}
\medskip

Projection operator on the spin $3/2$ state with mass $s^{1/2}$ and 
four-momenta $P=\Bigl(\sqrt{s+{\bf P}^2},{\bf P}\Bigr)$ is  
\cite{Nie,W,Bammer,Gas}

$$\Lambda^{a,b}({\bf P})=\sum_{S=-3/2}^{3/2}
u^{a}({\bf P},S){\overline u}^{b}({\bf P},S)=
{ {\gamma_{\sigma}P^{\sigma}+s^{1/2} }\over {2s^{1/2}} }
\biggl[{{\it P}^{3/2} }\biggr]^{ab}
\eqno(A.1a)$$

where 

$$\biggl[{{\it P}^{3/2} }\biggr]^{ab}=
g^{ab}- {1\over 3} \gamma^{a}\gamma^{b}-
{1\over {3s}}\bigl(\gamma_{\sigma}P^{\sigma}
\gamma^{a}P^{b}+
\gamma^{b}P^{a}
\gamma_{\sigma}P^{\sigma}\bigr)\eqno(A.2a)$$

These expressions were derived \cite{Fronsdal,W} imposing a restriction
$\gamma_a\biggl[{{\it P}^{3/2} }\biggr]^{ab}=0$ and using  the commutation 
condition of  $\biggl[{{\it P}^{3/2} }\biggr]^{ab}$ and 
${ {(\gamma_{\sigma}P^{\sigma}+s^{1/2}) }/ {(2s^{1/2})} }$ 
projection operators.
In consequence of  these conditions one obtains a result
an identical with  (A.1a)

$$\Lambda^{a,b}({\bf P})=\biggl[{{\it P}^{3/2} }\biggr]^{ab}
{ {\gamma_{\sigma}P^{\sigma}+s^{1/2} }\over {2s^{1/2}} }.\eqno(A.1b)$$

On mass shell $\biggl[{{\it P}^{3/2} }\biggr]^{ab}$
is equal to a  well known expression \cite{W}

$$\biggl[{{\it P}^{3/2} }\biggr]^{ab}=
g^{ab}- {1\over 3} \gamma^{a}\gamma^{b}-
{{2P^{a}P^{b}}\over {{3s} } } +
{{\gamma^{a}P^{b}-\gamma^{b}P^{a} }\over {3s^{1/2} }}
\eqno(A.2b)$$

Operator $\biggl[{\it P}^{3/2}\biggr]^{ab}$ (A.2a,b) together with 
spin $1/2$ projection operators $\biggl[{\it P}^{1/2}\biggr]^{ab}$
satisfy a completeness condition

$$\biggl[{{\it P}^{3/2} }\biggr]^{ab}+
\biggl[{{\it P}^{1/2}_{11}}\biggr]^{ab}+
\biggl[{{\it P}^{1/2}_{22}}\biggr]^{ab}=g^{ab},\eqno(A.3)$$

where

$$
\biggl[{{\it P}^{1/2}_{11}}\biggr]^{ab}=
 {1\over 3} \gamma^{a}\gamma^{b}
-{2\over {3s}}P^{a}P^{b}+{1\over {3s}}\bigl(\gamma_{\sigma}P^{\sigma}
\gamma^{a}P^{b}+\gamma^{b}P^{a}\gamma_{\sigma}P^{\sigma}\bigr),\eqno(A.4a)$$

$$
\biggl[{{\it P}^{1/2}_{22}}\biggr]^{ab}=
{2\over {3s}}P^{a}P^{b}
\eqno(A.4b)$$

Afterwards the completeness condition
for the projection operators of spin $3/2$ particles takes a form

$$
\sum_{S=-3/2}^{3/2}\Biggl(
 u^{a}({\bf P},S){\overline u}^{b}({\bf P},S)+
{ {\gamma_{\sigma}P^{\sigma}+s^{1/2} }\over {2s^{1/2}} }
\biggl\{\biggl[{{\it P}^{1/2}_{11}}\biggr]^{ab}+
\biggl[{{\it P}^{1/2}_{22}}\biggr]^{ab}\biggr\}\Biggr)+$$
$$\sum_{S=-3/2}^{3/2}\Biggl(
 v^{a}({\bf P},S){\overline v}^{b}({\bf P},S)+
{ -{\gamma_{\sigma}P^{\sigma}+s^{1/2} }\over {2s^{1/2}} }
\biggl\{\biggl[{{\it P}^{1/2}_{11}}\biggr]^{ab}+
\biggl[{{\it P}^{1/2}_{22}}\biggr]^{ab}\biggr\}^{\ast}\Biggr)
=g^{ab}\eqno(A.5)$$

where $v^{a}({\bf P},S)$ is bispinor of antiparticle with spin $3/2$.

Now one can decompose the $\gamma N-N$ vertex function over the intermediate 
spin $3/2$ states

$$
g_{bc}{\overline u}({\bf p'_N})\gamma_5g^{ab}
\Bigl[e_{N'}{(P+P')}^{\mu}-i\mu_{N'}\sigma^{\mu\nu}k'_{\nu}\Bigr]
g^{cd}\gamma_5 u({\bf p'_N+k'})=$$
$${\overline u}({\bf p'_N})\gamma_5
\sum_{S=-3/2}^{3/2}\Biggl(
 u^{a}({\bf P},S){\overline u}^{b}({\bf P},S)+
v^{a}({\bf P},S){\overline v}^{b}({\bf P},S)+...\Biggr)
g_{bc}\Bigl[e_{N'}{(P+P')}^{\mu}-i\mu_{N'}\sigma^{\mu\nu}k'_{\nu}\Bigr]$$
$$\sum_{S'=-3/2}^{3/2}\Biggl(
 u^{c}({\bf P},S){\overline u}^{d}({\bf P},S)+
v^{c}({\bf P},S){\overline v}^{d}({\bf P},S)+...\Biggr)
\gamma_5u({\bf p'_N+k'})\eqno(A.6a)$$

In the $\Delta$-resonance region one can take into account only
spin $3/2$ intermediate states. Then without 
antiparticle degrees of freedom and without spin $1/2$ intermediate states
we get

$$\biggl[(p'_{N})_{a}(p_N)^{a}{\overline u}({\bf p'_N})\gamma_5
\Bigl[e_{N'}{(P+P')}^{\mu}-i\mu_{N'}\sigma^{\mu\nu}k'_{\nu}\Bigr]\gamma_5
u({\bf p'_N+k'})\biggr]^{Projection\ on\ spin\ 3/2\ states}=$$
$$\sum_{S,S'=-3/2}^{3/2} 
{\overline u}({\bf p'_N})\gamma_5{p'_N}_au^a({\bf P'},S')\Biggl\{
{\overline u}^b({\bf P'},S')g_{bc}\biggl[e_{N'}{(P+P')}^{\mu}$$
$$-i\mu_{N'}\sigma^{\mu\nu}k'_{\nu}\biggr]u^c({\bf P},S)\Biggr\}
{\overline u}^d({\bf P},S)(p_N)_d\gamma_5
u({\bf p'_N+k'}). \eqno(A.6b)$$

The completeness condition (A.6a) with the spin $3/2$ states only 
enables us to 
represent the $\pi N$ amplitude in the convenient form
 
$$<out;{\bf p'}_{N}|j_{\pi'}(0)|{\bf p}_{\pi}{\bf p}_N;in>\Longrightarrow
{\overline u}({\bf p'}_N)\gamma_5^2 u({\bf p}_N)
{\overline u}({\bf p}_N)u({\bf p'}_N)
<out;{\bf p'}_{N}|j_{\pi'}(0)|{\bf p}_{\pi}{\bf p}_N;in>\Longrightarrow
{1\over{(p'_N{\bf .}p_N) }}$$
$$\Biggl[
{\overline u}({\bf p'}_N)\gamma_5(p'_N)_au^a({\bf P'})\biggl\{
{\overline u}^b({\bf P'})g_{bc}u^c({\bf P})\biggr\}
{\overline u}^d({\bf P})(p_N)_d\gamma_5u({\bf p}_N)\Biggr]
{\overline u}({\bf p}_N)u({\bf p'}_N)
<out;{\bf p'}_{N}|j_{\pi'}(0)|{\bf p}_{\pi}{\bf p}_N;in>
\eqno(A.7)$$

where for the sake of simplicity we have omitted the 
spin indices $S$ and $S'$.

\newpage

\vspace{0.25cm}

\begin{center}
   {{\bf Appendix B:} $\gamma'\Delta'-\Delta$ vertex function with on mass 
shell $\Delta$-s}

\end{center}

\vspace{0.25cm}

Relations (3.19a,b) for the $\Delta-\gamma\Delta$ vertex functions are 
obtained from the spin $3/2$ particle electromagnetic vertex function

$$<out;{\bf P'}|J^{\mu}(0)|{\bf P};in>=
=(P+P')^{\mu}\Biggl(
{\overline u}^{\sigma}({\bf P'})
\Bigl[g_{\rho\sigma}G_1({k'}^2)
+{k'}_{\sigma}{k' }_{\rho}G_3({k'}^2)
\Bigr]\Biggr)u^{\rho}({\bf P})$$
$$+{\overline u}^{\sigma}({\bf P'})\Biggl(
-i\sigma^{\mu\nu}{k'}_{\nu}\Bigl[g_{\rho\sigma}G_2({k'}^2)
+{k'}_{\sigma}{k' }_{\rho}G_4({k'}^2)
\Bigr]\Biggr)u^{\rho}({\bf P}),\eqno(B.1)$$

where ${k'}_{\mu}=(P-P')_{\mu}$ is the four-momentum of the emitted photon,
${\bf P}={\bf p}_N+{\bf p}_{\pi}=
{\bf P}_{\Delta}$, ${P}^o=\sqrt{m_{3/2}^2+{\bf P}^2}$,
 ${\bf P'}={\bf p}'_N+{\bf p'}_{\pi}={\bf P'}_{\Delta}$, 
 ${P'}^o=\sqrt{m_{3/2}^2+{\bf P'}^2}$ are the four-momentum of a spin $3/2$
particle with a mass $m_{3/2}$ in the initial and final states.

Extension in the complex mass and in the complex energy regions implies 
a transformations
$m_{3/2}\Longrightarrow m_{\Delta}=M_{\Delta}-i/2\Gamma_{\Delta}$ and 
$k'\Longrightarrow k'_{\Delta}$.
Here it is assumed that $M_{\Delta}=1232MeV or 1210MeV$.

 Functions $G_i({k'}^2)$ in (B.1) are real. Therefore 
an important property of the electromagnetic $\Delta$
 vertex functions (3.19a,b) is that the electromagnetic 
constants  and  $G_i(0)$ in (B.1) and in (3.19a,b) are also real.

The spin $3/2$ particle vertex function (B.1) 
satisfy the one-particle Ward-Takahashi identity
which in the complex energy region of $\Delta$-resonance has the form

$${k'_{\Delta}}^{\mu}<{\bf P'}_{\Delta}|J_{\mu}(0)|{\bf P}_{\Delta}>=
{\overline u}^{\sigma}({\bf P'}_{\Delta})
\biggl[ S^{-1}_{\sigma\rho}( {P'}_{\Delta}^o,{\bf P'}_{\Delta})
-S^{-1}_{\sigma\rho}( {P}_{\Delta}^o,{\bf P}_{\Delta})
\biggr]_{s_{\Delta}=s'_{\Delta}=m_{\Delta}^2}
u^{\rho}({\bf P}_{\Delta}),\eqno(B.2)$$

where $S_{\sigma\rho}( {P'}_{\Delta}^o,{\bf P'}_{\Delta})$
denotes a propagator of $\Delta$ \cite{Amiri,PasVan2,PasVan3}. 

It must be emphasized that extraction of the $\Delta$ degrees of freedom 
considered in this paper 
does not use a Heisenberg local field operator
of the $\Delta$ resonance or
a Lagrangian with the $\Delta$ degrees of 
freedom, because it is not possible to construct a Fock space for a ``free''
resonance state with a complex mass. In the present approach we operate only  
with vertex functions with on mass shell $\Delta$'s which are obtained 
in the limit $s\Longrightarrow m_{\Delta}^2$. 
Therefore  ambiguities generated 
by unphysical gauge transformations of the $\Delta$-particle field operator
$\Psi_{\Delta}^a\longrightarrow \Psi_{\Delta}^a+C\gamma^a\gamma_b 
\Psi_{\Delta}^b$ \cite{Bammer} with an arbitrary parameter $C$
does not appear in the present formulation.
A sensitivity of the $\gamma p\to 
\gamma'\pi' p'$ observable on the choice of the form of the intermediate 
$\Delta$ propagator is demonstrated in \cite{MF}.

In the off  mass shell region  one has four independent
momenta of each $\Delta$'s because
${P'}^2\ne m^2_{\Delta}$ and 
${P}^2\ne m^2_{\Delta}$. Therefore for the off mass shell $\Delta$'s
(3.19a,b) takes more complicated 
form with increasing number of  form factors $G_i$ because 
each of the conditions
$P^2_{\Delta}-m_{\Delta}^2\ne 0$ and  
$(i\gamma_{\sigma}P_{\Delta}^{\sigma}-m_{\Delta}^2)\ne 0$ 
doubling the number of form factors. Therefore instead 
of two form factors in (3.19b) we get $8$ form factors for the off mass shell
$\Delta-\gamma'\Delta'$  vertices. The role of these six 
additional formfactors is so important as important is the off mass shell 
terms in the $\Delta$ propagator $S_{\sigma\rho}$. 
Besides these formfactors of the $\Delta-\gamma'\Delta'$ vertex
with off mass shell $\Delta$'s are depending on three variables 
${k'}^2_{\Delta}$, $P^2_{\Delta}$ and  ${P'}^2_{\Delta}$.
Therefore use of the off 
mass shell $\Delta$ propagators together with the on mass shell 
$\Delta-\gamma'\Delta'$ as it is done in refs. \cite{Amiri,PasVan2,PasVan3}
is inconsistent.

Now we consider the recipe of extraction of the double $\Delta$ exchange
term (3.18a) from the $\pi N$ bremsstrahlung amplitude (2.6).
According to our previous paper \cite{Ann,MF}, we consider $s$-channel part
of the  Green function $\tau_{\mu}$ (2.2a)
with the  $\pi N$ $''in''$ asymptotic states

$${\tau}^{\mu}\Longrightarrow\sum_{ {\pi N},{\pi'N'}}
 <0|{\sf T}\biggl(\Bigl[\Psi(y')\Phi(x')\Bigr]|{\pi}'N';out>
<out;\pi'N'|{\cal J}^{\mu}(0)|{\pi N};in><in;{\pi N}|
\Bigl[{\overline \Psi}(y)\Phi^+(x)\Bigr]\biggr)|0>$$
$$\Longrightarrow
\sum_{ {\pi N},{\pi'N'}}<...{\cal G}_{\pi'N'} 
\Bigl[<out;\pi'N'|{\cal J}^{\mu}(0)|{\pi N};in>\Bigr]_{\pi\ N\ irreducible}
{\cal G}_{\pi N }...>,\eqno(B.3)$$
where ${\cal G}_{\pi N }$ denotes a full 
$\pi N$ Green function. This Green function can be decomposed over the full
system of the $\pi N$ wave functions $|\psi_{\pi N}>$ which afterwards 
can be replaced by the $\Delta$ wave function   $|\psi_{\Delta}>$

$${\cal G}_{\pi N }=\sum_{\pi N} 
{ {|{\widetilde \psi}_{\pi N}><\psi_{\pi N}| }\over{E-E_{\pi N}}}
\simeq\sum_{\Delta}
{ {|{\widetilde \psi}_{\Delta}><\psi_{\Delta}| }\over{E-E_{\Delta}}}.
\eqno(B.4)$$

Now if we take into account, that 
$<\Psi_{\Delta'}|{\cal J}^{\mu}(0)|{\widetilde \psi}_{\Delta}>$ is the
vertex function (3.19a) and $<...|\Psi_{\pi'N'}>$ 
produces expression (3.16a), then we obtain (3.19a) after
substitution of (B.4) into (B.3).

\newpage


\end{document}